\documentclass[letterpaper,superscriptaddress,english,aps,prb,twocolumn,floatfix, showpacs, amsfonts, amssymb]{revtex4}
\usepackage{epsfig, graphicx,psfrag,amsmath,amssymb,float}
\usepackage[T1]{fontenc}
\usepackage[latin9]{inputenc}
\usepackage{amsmath}
\usepackage{amssymb}
\usepackage{bbold}
\usepackage{amscd}
\usepackage{bm}
\usepackage{psfrag}

\usepackage{babel}

\newcommand{\be}{\begin{equation}}
\newcommand{\ee}{\end{equation}}
\newcommand{\bea}{\begin{eqnarray}}
\newcommand{\eea}{\end{eqnarray}}

\newcommand{\mc}{\mathcal}

\begin{document}

\title{Charge dynamics in half-filled Hubbard chains with finite on-site interaction}

\author{R. G. Pereira}
\affiliation{Instituto de F\'{i}sica de S\~ao Carlos, Universidade de S\~ao Paulo, C.P. 369, S\~ao Carlos, SP,  13560-970, Brazil}

\author{K. Penc}
\affiliation{Research Institute for Solid State Physics and Optics, H-1525 Budapest, P.O.B. 49, Hungary}

\author{S. R. White}
\affiliation{Department of Physics and Astronomy, University of California, Irvine, CA 92617, USA}

\author{P. D. Sacramento}
\affiliation{CFIF, Instituto Superior T\'{e}cnico, Universidade T\'{e}cnica de Lisboa, Av. Rovisco Pais, 1049-001 Lisboa, Portugal}

\author{J. M. P. Carmelo}
\affiliation{GCEP-Centre of Physics, University of Minho, Campus Gualtar, P-4710-057 Braga, Portugal}
\affiliation{Institut f\"ur Theoretische Physik III, Universit\"at Stuttgart, D-70550 Stuttgart, Germany}

\begin{abstract}
We study the charge dynamic structure factor of the one-dimensional Hubbard model with finite on-site repulsion $U$ at half filling. Numerical results from the time-dependent density matrix renormalization group are analyzed by comparison with the exact spectrum of the model. The evolution of the line shape as a function of $U$ is explained in terms of a relative  transfer of spectral weight between the two-holon continuum that dominates in the limit $U\to\infty$ and a subset of the two-holon-two-spinon continuum that reconstructs the electron-hole continuum in the limit $U\to0$.  Power-law singularities along boundary lines of the spectrum are described by  effective impurity models that are explicitly invariant under  spin and $\eta$-spin $SU(2)$ rotations. The Mott-Hubbard metal-insulator transition  is reflected in a discontinuous change of the exponents of edge singularities at $U=0$. The sharp feature observed in the   spectrum for momenta near the zone boundary is attributed to a Van Hove singularity that persists as a consequence of integrability.

\end{abstract}


\maketitle

\section{Introduction}

Since its proposal,\cite{hubbard} the Hubbard model has become a paradigm in the field of strongly correlated electron systems. It is the simplest model that accounts for the metal-insulator transition on a half-filled lattice when the on-site electron-electron repulsion $U$ is strong enough. It is still   debated whether the model in two spatial dimensions or some variation of it contains the mechanism for high temperature superconductivity at finite doping. 

Theoretically, much more is known about the model on a one-dimensional (1D) lattice.\cite{esslerbook} In this case, it is possible to calculate the exact spectrum and eigenfunctions by Bethe ansatz (BA).\cite{liebwu} Two remarkable properties revealed by  the exact solution are the existence of fractional excitations that carry separate spin and charge quantum numbers  and the opening of the Mott-Hubbard gap at half filling for arbitrarily small $U>0$. 

Recently, there has been renewed interest in dynamical properties of 1D models. One motivation for this is that  questions about features of the excitation   spectrum of 1D systems, such as the persistence of spin-charge separation at high energies, have become relevant with the improvement in the resolution of momentum-resolved experiments.\cite{gu,mzhasan,rixs,kim,jompol} In addition, ultracold atoms trapped in optical lattices have emerged as a new means to study coherent dynamics of 1D models, including integrable ones which are not realizable in condensed matter systems.\cite{bloch} 

At the same time,  significant progress has been achieved in developing analytical\cite{carmelo, pustilnik, cheianov, pereira08,imambsu2,imambekovscience,pereira10,essler10,sela,schmidt,schmidtprb,hamed, RMPimamb}   and numerical\cite{tdmrg,jeckelmann,kohno}  techniques to study dynamical correlation functions in the high energy regime where conventional  Luttinger liquid theory\cite{haldane, giamarchi} does not apply. Analytically, it is possible to compute  exponents of power-law singularities that develop near thresholds of the  spectrum of dynamical correlation functions at \emph {arbitrarily high energies}.  
 For the metallic phase of the Hubbard model, i.e. away from half filling, the calculation of finite-energy dynamical correlation functions  was pioneered by the pseudofermion dynamical theory.\cite{carmelo}
This theory is based on the BA solution of the model and has been applied to calculate, for instance, the optical conductivity and the one-electron spectral function  of 1D conductors.\cite{carmelo00,sing,lopes} In another  approach, exponents of high energy singularities can be investigated using effective field theories that treat high energy modes as impurities, defined in momentum space, which can scatter off low energy excitations (see Ref.  \onlinecite{RMPimamb} for a review). This approach, combined with the BA solution, has  also been applied to calculate edge exponents for  the spectral function of the Hubbard model away from half filling. \cite{essler10}

In this work we are interested in finite energy dynamical correlation functions for the Hubbard model at half-filling. Clearly, the edge exponents of the Mott insulating phase should differ from those of the metallic phase studied in Refs. \onlinecite{carmelo,essler10} due to the finite charge gap.  In fact, the result should be simpler due to the higher symmetry of the model at half filling. At half filling  the Hubbard model has a hidden $\eta$-spin $SU(2)$ symmetry that rotates between doubly occupied sites and empty sites.\cite{esslerbook}    In the same sense that spin $SU(2)$ invariance fixes the exponents of spin correlation functions  at high energies,\cite{imambsu2} it should be possible to use the continuous symmetry in the charge sector to constrain the  exponents in charge dynamics.

Particularly, we shall focus on the charge dynamic structure factor (DSF) $S(q,\omega)$  at zero temperature. The DSF is known analytically only in two limits. In the low energy limit, which requires that the  Mott-Hubbard gap be small, dynamical correlation functions can be calculated using form factors for the integrable sine-Gordon model.\cite{controzzi} Within this field theory approach, a universal square-root cusp is found at the edge of the relativistic spectrum of massive charge solitons. In the strong coupling limit, one can take advantage of the  factorization of the wave function into a noninteracting charge sector and a spin sector described by the 1D Heisenberg model.\cite{ogatashiba}    A square-root cusp is again found at the lower threshold of the two-holon continuum, stemming from matrix element for noninteracting spinless fermions.\cite{penc00} In both low energy and strong coupling limits, no features are predicted at the branch line of the spin excitations here called spinons. This is in contrast to the behavior of  the one-electron spectral function, which has sharp features near both charge and spin branch lines for any value of $U>0$.\cite{sorella,voit} 

The DSF has also been studied numerically,\cite{bannister,preuss} most recently using the dynamical density matrix renormalization group (DDMRG).\cite{jeckelmann} Most of the numerical work has focused on the regime of large $U$, which is appropriate to describe strong Mott insulators such as Sr$_2$CuO$_3$.\cite{mzhasan} 

We note that the DSF has strikingly different line shapes in the weak and strong coupling limits. For noninteracting electrons, $U=0$, the DSF can be calculated exactly and corresponds to the density of states of an electron-hole pair.  For $U\to\infty$, the spectral weight is assigned to a two-holon continuum, with negligible contribution from spinons.\cite{penc00} 

The purpose of this paper is to investigate the charge DSF for the Hubbard model at half filling  for arbitrary values of $q$, $\omega$  at \emph{finite  $U$}. We  construct a picture for the intermediate $U$ regime by combining  information about the exact spectrum from  BA, an  effective field theory for  edge singularities at high energies and numerical results from the time-dependent density matrix renormalization group (tDMRG). We start in Sec. \ref{sec:model} by  discussing the exact support of the DSF in terms of elementary charge and spin excitations using  known results from the BA solution. Our main results can be found in sections \ref{sec:edgetheory} and \ref{sec:DMRG}. In  Sec. \ref{sec:edgetheory} we present the effective field theory that incorporates the spin and $\eta$-spin $SU(2)$ symmetries explicitly and allow  us to determine the exponents of   power-law singularities  at the  edges of the exact spectrum of the DSF. In Sec. \ref{sec:DMRG} we present  the tDMRG results for certain values of $U$ and analyze them by comparison with the field theory combined with the exact spectrum from BA. In addition, we discuss the $U$ dependence of the line shape, interpolating between weak and strong coupling limits. Finally, Sec. \ref{sec:disc} contains the conclusions.

Our results  are relevant for the charge DSF of fermionic atoms in a 1D optical lattice with on-site atomic repulsion   described by the integrable Hubbard model.  In the context of cold atoms, the charge DSF is probed by Bragg spectroscopy.\cite{rey,ernst} The results are also useful as an approximation to condensed matter systems where the integrability-breaking perturbations to the Hubbard model, such as the nearest neighbour interaction in the extended Hubbard model, are small. In this context the DSF has served to interpret electron energy loss spectroscopy \cite{gu} and inelastic x-ray experiments.\cite{mzhasan,rixs}   Since the Shiba transformation\cite{shiba} maps the charge DSF for $U>0$  to the spin DSF for $U<0$, our results also apply to the spin DSF of the spin-gapped phase for the attractive Hubbard model. 

\section{Model, symmetries and exact support of the charge DSF \label{sec:model}}

\subsection{Model}
We consider the 1D Hubbard model \be\label{Hubbard}
H=\sum_{j=1}^L\left[-(c^\dagger_{j}c^{\phantom\dagger }_{j+1}+\textrm{h.c.})+U\left(n_{j,\uparrow}-\frac12\right)\left(n_{j,\downarrow}-\frac12\right)\right].
\ee
Here $c_j=(c_{j,\uparrow},c_{j,\downarrow})$ is a two-component spinor representing electrons with  spin $\sigma=\uparrow,\downarrow$ at site $j$,  $n_{j,\sigma}=c^\dagger_{j,\sigma}c^{\phantom\dagger}_{j,\sigma}$, and $L$ is the system size.  The number of electrons at site $j$ is denoted as $n_j=n_{j,\uparrow}+n_{j,\downarrow}$. We focus on half filling $\langle n_j\rangle =1$.  We have set the hopping amplitude to 1, not to be confused with the real time variable $t$. The 1D Hubbard model is integrable. The exact  spectrum for all values of $U$, density and magnetization is provided by the BA solution.\cite{liebwu}

At half filling and zero magnetic field, the Hubbard model has   an explicit  spin $SU(2)$ symmetry and a less obvious charge $\eta$-spin  $SU(2)$ symmetry.\cite{esslerbook} The generators of spin rotations are the components of the usual spin operator $\mathbf{S}=\sum_jc^\dagger_j(\bm{\tau}/2)c^{\phantom\dagger}_j$, where $\bm{\tau}$ is the vector of Pauli matrices. The generators of    $\eta$-spin rotations are  \bea
\eta^z&=&\frac12\sum_j(n_{j,\uparrow}+n_{j,\downarrow}-1)\equiv \sum_j \eta_{j}^z,\\
\eta^+&=&\sum_j(-1)^jc_{j,\uparrow}^\dagger c_{j,\downarrow}^\dagger \equiv\sum_j \eta_{j}^+,\\
\eta^-&=&\sum_j(-1)^jc_{j,\downarrow}^{\phantom\dagger} c_{j,\uparrow}^{\phantom\dagger}\equiv \sum_j \eta_{j}^-,
\eea
such that the local operators obey the algebra $[\eta_j^z,\eta_{j^\prime}^{\pm}]=\pm\delta_{jj^\prime}\eta_j^\pm$ and $[\eta_j^+,\eta_{j^\prime}^{-}]=2\delta_{jj^\prime}\eta_j^z$. Notice that  $\eta_j^z$ is proportional to the fluctuation of the local charge density operator: $n_j=1+2\eta^z_j$. The transverse components $\eta^{x,y}$ of the   $\eta$-spin vector $\bm{\eta}=(\eta^x,\eta^y,\eta^z)$ can be defined by $\eta^\pm=\eta^x\pm i \eta^y$. 

While the spin and $\eta$-spin symmetries account for an $SO(4) = [SU(2)\times SU(2)]/Z_2$ symmetry, the global symmetry of the  Hubbard model was recently found to be larger and given by $[SO(4)\times U(1)]/Z_2=[SU(2)\times SU(2)\times U(1)]/Z_2^2$.\cite{ostlund} In addition, the 1D model has an infinite number of local conserved quantities associated with integrability. 

\subsection{Charge structure factor at half filling}

The charge DSF is defined as the Fourier transform of the density-density correlation function\bea\label{DSF}
S(q,\omega)&=&\frac{2\pi}{L}\sum_{\nu\neq GS}|\langle GS|n_q|\nu\rangle|^2\delta(\omega -E_\nu+E_{GS}) \nonumber\\
&=&4\sum_j e^{-iqj}\int_{-\infty}^\infty dt\, e^{i\omega t} \langle \eta^z_j(t) \eta^z_0(0)\rangle,\label{eq:Sqw}
\eea
where $n_q\equiv\sum_je^{-iqj}n_j$, $|GS\rangle$ is the ground state and $|\nu\rangle$ is an excited state with energy $E_\nu$. Since $S(q,\omega)=S(-q,\omega)$, in the following we set $q>0$ without loss of generality.
The DSF obeys the sum rule\be
\int_{-\pi}^\pi \frac{dq}{2\pi}\int_0^\infty \frac{d\omega}{2\pi}\,S(q,\omega)=2\langle n_{j,\uparrow}n_{j,\downarrow}\rangle\equiv 2\mathfrak D,\label{sumrule}
\ee
with the density of doubly occupied sites given exactly by\cite{joseandpedro}\be
 \mathfrak D=\int_0^\infty d\omega\frac{J_0(\omega)J_1(\omega)}{1+\cosh(\omega U/2)}\geq0.
\ee
For $U\to 0$, we have $ \mathfrak D\to 1/4$; for $U\to \infty$, the integrated spectral weight  vanishes as $ \mathfrak D\to \ln 2(2t/U)^2$.

The $\eta$-spin symmetry can be used to relate the DSF at half filling  to the correlation function for the pairing operators, which create doubly occupied or empty sites. The ground state for an even number of   sites is unique and is a singlet of both spin and $\eta$-spin rotations (quantum numbers $S=S^z=\eta=\eta^z=0$). Eq. (\ref{eq:Sqw}) can be rewritten as\be
S(q,\omega)=8\pi L\sum_{\nu\neq GS}|\langle GS|\eta^z_0|\nu\rangle|^2\delta_{q,P_\nu}\delta(\omega -E_\nu+E_{GS}), 
\ee
where $P_\nu$ is the lattice momentum of the eigenstate $|\nu\rangle$. By employing, for instance, the unitary transformation that rotates the $\eta$-spin vector by $\pi/2$ about the $y$ axis, $U=e^{-i\frac\pi2\eta^y}$, we can rewrite the matrix element\be
\langle GS|\eta^z_0|\nu\rangle=\langle GS|U^\dagger U\eta^z_0U^\dagger U|\nu\rangle=\langle GS|\eta^x_0 |\nu^\prime\rangle,
\ee
where $|\nu^\prime\rangle=U|\nu\rangle$ is also an eigenstate of $H$ with energy $E_{\nu^\prime}=E_{\nu}$, but with momentum $P_{\nu^\prime}=P_{\nu}+\pi$. The momentum shift  follows from the fact that   
 the lattice translation operator anticommutes with   $\eta^\pm$.\cite{esslerbook} We then have \bea
S(q,\omega)&=&8\pi L\sum_{\nu\neq GS}|\langle GS|\eta^x_0|\nu\rangle|^2\delta_{q,P_\nu+\pi}\nonumber\\
&&\times\delta(\omega -E_\nu+E_{GS})\\
&=&4\sum_j e^{-i(q+\pi)j}\int_{-\infty}^\infty dt\, e^{i\omega t} \langle \eta^x_j(t) \eta^x_0(0)\rangle, \nonumber
\eea
and likewise for the correlation function for $\eta^y_j$. Thus $S(q,\omega)$ can be viewed  as the longitudinal  component of the charge DSF tensor\be
S_c^{ab}(q,\omega)=4\sum_j e^{-iqj}\int_{-\infty}^\infty dt\, e^{i\omega t} \langle \tilde\eta^a_j(t) \tilde\eta^b_0(0)\rangle,
\ee
where $a,b=x,y,z$  and $\tilde \eta_j^z= \eta_j^z$, $ \tilde\eta_j^{x,y}=(-1)^j \eta_j^{x,y}$. In this notation,  $S(q,\omega)=S_c^{zz}(q,\omega)$. The $\eta$-spin $SU(2)$ symmetry implies\bea
S(q,\omega)&=&S_c^{+-}(q,\omega)/2\nonumber\\
&=&2\sum_j e^{-i(q+\pi)j}\int_{-\infty}^\infty dt\, e^{i\omega t} \langle \eta^+_j(t) \eta^-_0(0)\rangle.\eea
Therefore, up to the shift of total momentum by $\pi$, the line shape of the charge DSF is identical to that of the correlation function for pairing operators $\eta_j^\pm$. We can also write\be
S(q,\omega)=\frac43\sum_j e^{-iqj}\int_{-\infty}^\infty dt\, e^{i\omega t} \langle \tilde{ \bm\eta}_j(t)\cdot \tilde{\bm\eta}_0(0)\rangle.\label{Sqwinvariant}
\ee

For later reference, we mention that for $U=0$ the charge DSF in Eq. (\ref{eq:Sqw}) reduces to the density of states for excitations with a single electron-hole pair\be
S_0(q,\omega)=\frac{4\theta(\omega-\omega_-(q))\theta(\omega_+(q)-\omega)}{\sqrt{\omega_+(q)^2-\omega^2}},\label{freefermion}
\ee
where $\omega_-(q)=2\sin q$ and $\omega_+(q)=4\sin(q/2)$ are the lower and upper thresholds of the electron-hole continuum, respectively, and $\theta(\omega)$ is the Heaviside step function. Up to a factor of 2, this is the same result as for spinless fermions at half filling. The free electron  DSF has a step discontinuity at the lower edge and a square-root divergence at the upper edge, which stems from the Van Hove singularity of an electron and a hole with the same velocity.

\subsection{Elementary excitations in the Bethe ansatz solution}

In this subsection we review some BA  results  for the exact spectrum which will be useful for comparison with numerical results in section IV.
 
According to the BA solution,\cite{liebwu} the eigenstates of the 1D Hubbard model   can be constructed from elementary charge, $\eta$-spin   and spin  excitations. In the half filling case, it suffices to consider two branches of excitations, one in the charge  sector, which we call holons, and one in the spin   sector, which we call  spinons. (For the relation between the holons and spinons used here and the notation used e.g. in Refs. \onlinecite{ostlund,companion}, see Appendix A.)

\begin{figure}
\begin{center}
\includegraphics*[width=60mm]{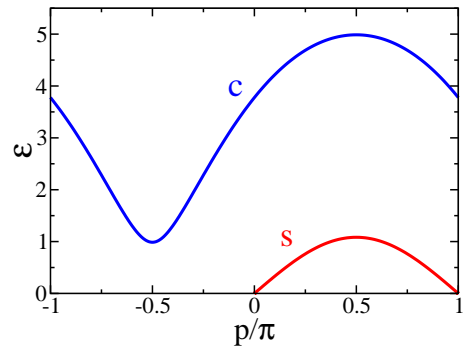}
\caption{(Color online.) Exact dispersion relations of elementary charge  and spin  excitations for $U=4.9$. Here the excitations are represented by particles in the empty  holon ($c$) and spinon ($s$) bands. The holon band  is gapped with minimum energy  at $p_c=-\pi/2$. The spinon band is gapless at $p_s=0,\pi$.
\label{fig:holonspinon}}
\end{center}
\end{figure}

   In the thermodynamic limit holons and spinons have dispersion relations $\varepsilon_{c}(p_{c})$ and $\varepsilon_s(p_s)$, respectively,  where the dressed momenta $p_{c,s}$ and dressed energies $\varepsilon_{c,s}$ are given by 
[see    Ref. \onlinecite{Carmelo92} and Appendix A; here we follow the notation in Eq. (7.8)  of Ref. \onlinecite{esslerbook}]   \bea
p_c(k)&=&\frac\pi2-k-2\int_0^\infty d\omega \frac{J_0(\omega)\sin(\omega\sin k)}{\omega (1+e^{\omega U/2})},\label{holonq}\\
p_s(\Lambda)&=&\frac\pi2-\int_0^\infty d\omega \frac{J_0(\omega)\sin(\omega \Lambda)}{\omega \cosh(\omega U/4)},\label{spinonq}\\
\varepsilon_c(k)&=&2\cos k+U/2\nonumber\\
&&+4\int^{\infty}_{0}d\omega \frac{J_1(\omega)\cos(\omega \sin k)}{\omega (1+e^{\omega U/2})},\label{holoneps}\\
\varepsilon_s(\Lambda)&=&2\int_0^{\infty}d\omega \frac{J_1(\omega)\cos(\omega \Lambda)}{\omega \cosh(\omega U/4)}.\label{spinoneps}
\eea
Here $k$ and $\Lambda$ are the charge quasimomentum and spin rapidity, respectively. 
For any value of $U>0$, the spin dispersion  is gapless at the spinon Fermi points $p_s=0,\pi$ and the charge dispersion has  minimum energy at $p_c= -\pi/2$,  with a gap given by\be
\Delta=\frac{16}{U}\int_1^\infty d\omega \frac{\sqrt{\omega^2-1}}{\sinh(2\pi \omega/U)}.
\ee
Analytic expressions for the holon and spinon dispersions can be obtained in the limits $U\to\infty$:\bea
\varepsilon_c(p_c)&\approx& \frac U2+2\sin p_c -\frac{4\ln2}{U} (1+\cos2p_c),\label{dispclargeU}\\
\varepsilon_s(p_s)&\approx&\frac4U \sin p_s,
\eea
and in the limit $U\to0$:\bea
\varepsilon_c(p_c)&\approx&4 \left|\cos \frac{2p_c-\pi}{4}\right|  ,\\
\varepsilon_s(p_s)&\approx& 2\sin p_s,\label{ecsmallU}
\eea
with an exponentially small charge gap $\Delta\approx (4/\pi)\sqrt{U}e^{-2\pi/U}$.
We shall also be interested in the velocity of holons and spinons, defined as \be
u_c(p_c)=\frac{\partial \varepsilon_c}{\partial p_c}, \qquad u_s(p_s)=\frac{\partial \varepsilon_s}{\partial p_s}.\ee
The dispersion relations of holons and spinons given by Eqs. (\ref{holonq} - \ref{spinoneps}) are illustrated in Fig. \ref{fig:holonspinon}. 

\subsection{Boundary lines in the exact spectrum of $S(q,\omega)$}

Even though the exact spectrum and wave functions of the 1D Hubbard model are known, it has not been possible to calculate the DSF directly from the BA solution. The difficulty is in computing the matrix elements in Eq. (\ref{eq:Sqw}) for significantly large chains. Unfortunately, unlike the Heisenberg model, there are so far no determinant formulas\cite{kitanine}  or vertex operator approach\cite{caux11} to compute form factors for the Hubbard model. 

Nonetheless, we can use the BA equations to compute the exact support of the DSF. It follows from the Wigner-Eckart theorem that the  excited states that contribute to $S(q,\omega)$ in Eq. (\ref{eq:Sqw}) must carry quantum numbers $S=S^z=0$ (spin singlets) and $\eta=1,\eta^z=0$ ($\eta$-spin triplets). This selects states with $2m$ holons, $m\geq1$, and $2n$ spinons, $n\geq0$. Since the excited states must contain at least two holons and the holon dispersion is gapped,  the DSF vanishes for $\omega<2\Delta$.

The simplest excited states, in the sense of lowest number of elementary excitations, that contribute to $S(q,\omega)$   are  two-holon states   ($m=1$, $n=0$). For $\eta=1$, $\eta^z=0$, the  excitations with $m=1$, $n=0$ have total momentum $P$ and energy $E$ given by\cite{esslerbook}\be
P=p_{c,1}+p_{c,2}+\pi,\quad E= \varepsilon_{c}(p_{c,1})+\varepsilon_{c}(p_{c,2}),\label{PEtwoholons}
\ee
where $p_{c,1}$ and $p_{c,2}$ are the dressed momenta of the individual holons as in Eq. (\ref{holonq}). The next simplest excited states that contribute to $S(q,\omega)$ contain two spinons in addition to the two holons  ($m=n=1$).  For $\eta=1$, $\eta^z=0$  excitations with $m=n=1$, we have\bea
P&=&p_{c,1}+p_{c,2}+p_{s,1}+p_{s,2}+\pi, \\
E&=& \varepsilon_{c}(p_{c,1})+\varepsilon_{c}(p_{c,2})+ \varepsilon_{s}(p_{s,1})+\varepsilon_{s}(p_{s,2}),
\eea
where $p_{s,1}$ and $p_{s,2}$ are the momenta of the two spinons as in Eq. (\ref{spinonq}). We expect these two classes of states to give the leading contributions to the spectral weight of $S(q,\omega)$ for all values of $U$, based on the observation that, analogously, the leading contribution to  the half-filling one-electron excitations stem from one-holon-one-spinon
excited states.\cite{karlojose}  Indeed, figure 2 of Ref. \onlinecite{karlojose}   presents the contributions of different states to the one-electron-addition
sum rule for half filling. Interestingly, the higher-order
contributions are most important at $U\approx 4$, yet they account only
for about $0.005$ of the one-electron-addition spectral weight.
Consistently, it is expected that the higher-order contributions
associated here mainly with $m=1,n=2$ and $m=2,n=0$ states
are again very small and maximum at $U\approx 4$.

\begin{figure}
\begin{center}
\includegraphics*[width=70mm]{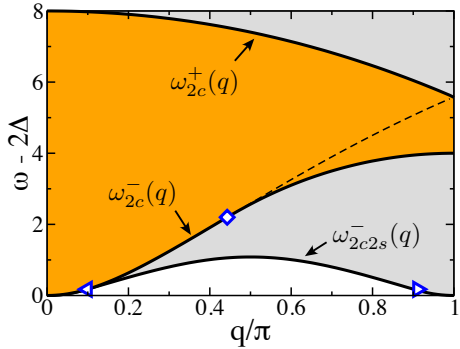}
\caption{(Color online.) Support of the charge DSF for  $U=4.9$.  Energies are measured from the Mott-Hubbard gap $2\Delta$. Some special lines of the spectrum are shown. The two-holon continuum is bounded by $\omega_{2c}^-(q)$ and $\omega_{2c}^+(q)$, but the spectral weight is nonzero everywhere above the lower line $\omega_{2c2s}^-(q)$ and extends to arbitrarily high energies. For $0<q<q_\diamond$, where $q_\diamond$ is the momentum of the point indicated by a diamond, the lower edge of the two-holon continuum $\omega_{2c}^-(q)$ is defined by two holons with the same momentum $(-\pi+q)/2$.  For $q_\diamond<q<\pi$, the energy of two holons with momentum ${(-\pi+q)/2}$ follows the dashed line, but this is no longer the lower edge of the two-holon spectrum. Instead, $\omega_{2c}^-(q)$ is defined by two holons with different momenta but equal velocities. For $0<q<q_\triangleleft$, where $q_\triangleleft$ is the momentum of the point indicated by the left-pointing triangle,  the DSF vanishes below $\omega_{2c}^-(q)$. For $q_\triangleleft<q<q_\triangleright\equiv\pi-q_\triangleleft$, the absolute lower threshold is $\omega_{2c2s}^-(q)$, defined by  an excitation with two holons with the same momentum, one spinon at the Fermi surface and another spinon below the Fermi surface that has the same velocity as the holons. For $q_\triangleright<q<\pi$, the line $\omega_{2c2s}^-(q)$ is defined by two holons with the same momentum and two spinons at opposite Fermi points. 
\label{continuum49}}
\end{center}
\end{figure}

Fig. \ref{continuum49} illustrates the exact support of the DSF. It also indicates special boundary lines in the $m=1,n=0$ and $m=n=1$ continua which will be important to construct the effective   field theory for edge singularities in Sec. \ref{sec:edgetheory} as well as to analyze the tDMRG data in Sec. \ref{sec:DMRG}.

We now discuss the most important boundary lines in the spectrum of $S(q,\omega)$  based on simple kinematics.
For large $U$,  we expect the spectral weight of $S(q,\omega)$ to  be confined inside the two-holon continuum.\cite{penc00}  The upper threshold of the two-holon continuum $\omega_{2c}^+(q)$ is given by two holons with the same momentum $(\pi+q)/2$, $0\leq q\leq\pi$.  In  the strong coupling theory for $U\to\infty$,\cite{penc00} in which limit the holons have a free-fermion cosine dispersion, the lower threshold of the two-holon continuum is given by two holons with the same momentum $p_{c,1}=p_{c,2}=(-\pi+q)/2$ for all $0\leq q\leq\pi$. However, for any finite $U$ the holon dispersion deviates from the cosine function such that the curvature of the dispersion (absolute value of inverse effective mass) is smaller near the minimum of the band than near the maximum. As a result, for values of $q$ near the zone boundary  the two-holon excitation with the lowest energy has holons with different momenta $p_{c,1}=q-\pi-p_{c,2}\neq p_{c,2}$ (mod $2\pi$), but such that they propagate with the same velocity, $u_c(p_{c,1})=u_c(p_{c,2})$. Starting from $p_{c,1}=p_{c,2}=-\pi/2$ and increasing the holon momenta, the values of $p_{c,1}$ and $p_{c,2}$ that define $\omega_{2c}^-(q)$ split off at the inflection point of the exact holon dispersion. Thus  there is a value   of $q_\diamond(U)$,  given by twice the momentum of the inflection point (plus or minus $\pi$ as in Eq. (\ref{PEtwoholons})), where the nature of the lower threshold changes. In the limit $U\to\infty$, Eq. (\ref{dispclargeU}) yields $q_\diamond\approx \pi-16\ln2/U+\mc{O}(U^{-2})$. Using the exact holon dispersion in Eqs. (\ref{holonq}) and (\ref{holoneps}) we find that $q_\diamond$ decreases monotonically with $U$ and $q_\diamond\to0$ in the limit $U\to0$.

The lower edge of the two holon continuum is not the absolute lower threshold of the support of $S(q,\omega)$ for general $q$. Starting from $q=0$ and moving along the line  $\omega_{2c}^-(q)$,   a value of $q$ is reached at which the velocity of the holons with momentum $(-\pi+q)/2$ becomes equal to the spin velocity at the spinon Fermi surface. The value of $q=q_\triangleleft(U)$ where this happens is given by the condition $u_c((-\pi+q_\triangleleft)/2)=v_s\equiv u_s(0)$ and is represented by a left-pointing triangle in Fig. \ref{continuum49}. For $q>q_\triangleleft$, it is possible to lower the energy by transferring momentum to a pair of spinons. For $q_\triangleleft<q<\pi-q_\triangleleft\equiv q_\triangleright$, the lower edge of the two-holon-two-spinon continuum, denoted $\omega_{2c2s}^-(q)$, has two holons with $p_{c,1}=p_{c,2}<(-\pi+q)/2$, one spinon at the Fermi point with $p_{s,1}=0$ and another spinon with momentum $p_{s,2}=q-2p_{c,1}$ such that the velocity of the latter equals the velocity of the holons, $u_s(p_{s,2})=u_c(p_{c,1})$. For $q_\triangleright<q<\pi$, the lower edge has the two spinons pinned at opposite Fermi points while the holons carry the same momentum $q/2$. 
 
The line $\omega_{2c2s}^-(q)$ is actually  the absolute lower edge of the support for $q_\triangleleft<q<\pi$. Adding more holons to the excited state can only increase the energy due to the charge gap. Furthermore, we find numerically that the spinon band has no inflection points away from the Fermi surface. In this case the minimum energy for $2n$ spinons at fixed total momentum is obtained by placing $2n-1$ spinons at the Fermi surface and one spinon carrying the remaining momentum, giving the same minimum energy as for two spinons only. Notice that the $\omega_{2c2s}^-(q)$ line is \emph{not} the same as the spinon mass shell, in contrast with the lower edge for the metallic phase.\cite{essler10,schmidtprb}

 Finally, we note that in the limit $U\to0$ the line $\omega_{2c2s}^-(q)$ becomes the lower edge of the electron-hole continuum, $\omega_{2c2s}^-(q)\to 2\sin q$, whereas the lower edge of the two-holon continuum becomes the \emph{upper} edge of the electron-hole continuum, $\omega_{2c}^-(q)\to 4\sin(q/2)$. As $U\to0$, we expect that all the spectral weight of $S(q,\omega)$ becomes confined between $\omega_{2c2s}^-(q)$ and $\omega_{2c}^-(q)$ in order to recover the free electron result.


\section{$SU(2)$ invariant impurity model for edge singularities\label{sec:edgetheory}}

In this section we work out the   field theory methods that allow us to describe power-law  singularities of dynamical correlation functions at high energies.  The general method  relies on effective impurity models  to treat the high energy modes. This approach has  been applied to other models and is explained in detail in Ref. \onlinecite{RMPimamb}. Here our goal is to extend these methods to incorporate  the spin and $\eta$-spin $SU(2)$ symmetries of the Hubbard model at half filling explicitly in the effective impurity models. The main idea is to define vector currents for the high energy modes, in analogy with the  low energy $SU(2)$ currents used in the Sugawara representation of the spin part of the Luttinger model.\cite{tsvelik}

\subsection{Low energy theory}

Before   dealing with high energy singularities, we review standard results obtained by bosonization of the Hubbard model  in the low energy limit.\cite{giamarchi} The starting point is to linearize the electron dispersion for $U=0$ about the right ($R$) and left ($L$) Fermi points for  the two spin channels  $\sigma=\uparrow, \downarrow$. In the continuum limit, the fermionic field is expanded in the form\be
c_{j,\sigma}\to \Psi_{\sigma}(x)\sim e^{i\pi x/2}\psi_{R,\sigma}(x)+e^{-i\pi x/2}\psi_{L,\sigma}(x).
\ee  
Bosonization maps the fermionic fields to\be
\psi_{\alpha,\sigma}(x)\sim F_{\alpha,\sigma} e^{-i\sqrt{2\pi}\varphi_{\alpha,\sigma}(x)},
\ee
for   $\alpha=L,R=+,-$, where $F_{\alpha,\sigma}$ are Klein factors. The chiral bosonic fields satisfy $[\varphi_{\alpha,\nu}(x),\partial_{x^\prime}\varphi_{\alpha^\prime,\nu^\prime}(x^\prime)]= i\alpha\delta_{\alpha,\alpha^\prime}\delta_{\nu,\nu^\prime}\delta(x-x^\prime)$. Charge and spin bosons are defined as the linear combinations\bea
\varphi_{\alpha,c}(x)=&[\varphi_{\alpha,\uparrow}(x)+\varphi_{\alpha,\downarrow}(x)]/\sqrt2,\\
\varphi_{\alpha,s}(x)=&[\varphi_{\alpha,\uparrow}(x)-\varphi_{\alpha,\downarrow}(x)]/\sqrt2.
\eea
The long wavelength part of the  spin and $\eta$-spin density operators can be expressed in terms of the chiral spin and charge bosons as \bea
\mathbf S_j &\to& \mathbf S(x)\sim \mathbf{J}_{R,s}(x)+ \mathbf{J}_{L,s}(x), \label{Sfield}\\
\bm\eta_j &\to& \bm\eta(x)\sim \mathbf{J}_{R,c}(x)+ \mathbf{J}_{L,c}(x), \label{etafield}
\eea
where $\mathbf{J}_{\alpha,\nu}$ with $\nu=c,s$ are $SU(2)$ charge and spin currents with components \bea
 J_{\alpha,\nu}^z(x)&=&\alpha\partial_x \varphi_{\alpha,\nu}(x)/\sqrt{4\pi},  \\
 J^\pm_{\alpha,\nu}(x)&=&e^{\pm i\sqrt{4\pi}\varphi_{\alpha,\nu}(x)}/2\pi.
\eea
These $SU(2)$ currents obey the $k=1$ Kac-Moody algebra.\cite{tsvelik} We remark that the long wavelength parts of $\mathbf S(x)$ and $\bm \eta(x)$ do  not mix charge and spin bosons, but  the staggered parts  omitted in Eqs. (\ref{Sfield}) and (\ref{etafield}) do.\cite{tsvelik}

In the low-energy limit, spin-charge separation holds in the strong sense that spin and charge excitations are decoupled.
The bosonized version of the Hubbard model in Eq. (\ref{Hubbard}) yields the Hamiltonian density $\mathcal{H}(x)=\sum_{\nu=c,s}[\mathcal{H}^{(0)}_\nu(x)+\delta \mathcal{H}_\nu(x)]$ with \bea
\mathcal{H}^{(0)}_\nu&=&\frac{2\pi v_\nu}3(\mathbf{J}_{R,\nu}^2+\mathbf{J}_{L,\nu}^2),\label{hspin}\\
\delta \mathcal{H}_\nu&=&-2\pi v_\nu \lambda_\nu\mathbf{J}_{R,\nu}\cdot \mathbf{J}_{L,\nu}. 
\eea
The terms $\mathcal{H}^{(0)}_\nu$  are quadratic in the bosonic fields and can be recognized as the Luttinger model for charge and spin collective modes written in manifestly $SU(2)\times SU(2)$ invariant form.  The parameters $v_c $ and $v_s$ are the charge and spin velocities, respectively. For $U\ll1$, we have $v_c\approx 2+U/2\pi$ and $ v_s\approx 2-U/2\pi$. The terms $\delta\mathcal{H}_\nu$  are perturbations that mix $R$ and $L$ currents and are not quadratic in the bosonic fields. For $U\ll1$, $\lambda_c\approx -U/2\pi$ and $\lambda_s\approx U/2\pi$. Although the bare coupling constants $\lambda_\nu$ are small for $U\ll1$, these perturbations flow under   the renormalization group  with  $\beta$ function\be
\frac{d\lambda_\nu}{d\ell}=-\lambda_\nu^2+O(\lambda_\nu^3),
\ee
where $d\ell=|d \Lambda|/\Lambda$ with $\Lambda$ the high energy cutoff. For $U>0$, $\lambda_s$ is marginally irrelevant and the spin spectrum is gapless. On the other hand, $\lambda_c$ is marginally relevant  and gives rise to a charge gap. The gap $\Delta\sim e^{-1/|\lambda_c|}$ is exponentially small  at small $U$, in agreement with the BA solution (c.f. below Eq. (\ref{ecsmallU})). The charge sector can then be described using the sine-Gordon model,\cite{controzzi}  whose elementary excitations are solitons with a massive relativistic dispersion $\epsilon(q)=\sqrt{(v_cq)^2+\Delta^2}$. Note the roles of spin and charge bosons are exchanged if we invert the sign of $U$,  as follows from  the Shiba transformation.\cite{shiba}

The critical theory of the spin sector is the $k=1$ $SU(2)$ Wess-Zumino-Witten (WZW) model.\cite{affleck} In the  more elegant notation of non-Abelian bosonization,  operators  can be written in terms of the $2\times2$ unitary matrix field $g(x)$ of the WZW model, \be
g(x,t)= \frac1{\sqrt2}g^{\phantom\dagger}_L(x_+)\otimes g^\dagger_R( x_-),
\ee
where   $x_\pm\equiv v_st\pm x$ and the tensor product notation means $g_{i,j}=g^{\phantom\dagger}_{L,i}g^\dagger_{R,j}$ with $i,j=1,2$. The chiral spinor fields $g_L$ and $g_R$ have conformal dimensions $(\frac14,0)$ and $(0,\frac14)$,\cite{bigyellowbook}   respectively, and can be represented in  Abelian bosonization  notation as \be
g_\alpha(x)= \left( {\begin{array}{c}
 e^{-i\sqrt{\pi}\varphi_{\alpha,s}(x)} \\
 e^{i\sqrt{\pi}\varphi_{\alpha,s}(x)}  \\
 \end{array} } \right),\qquad\alpha=L,R.\label{galpha}
\ee
Under a spin rotation represented by a unitary $2\times2$ matrix $U$, the chiral spinors transform as $g_{\alpha,_i}\to g_{\alpha,i} ^\prime=U_{ij}g_{\alpha,j}$. 
Due to conformal invariance,  the spin $SU(2)$ symmetry is enlarged to a chiral $SU(2)_L\times SU(2)_R$ symmetry. In terms of the matrix field, the spin currents are given by\cite{affleck}\be
\mathbf J_{L,s}=\frac{i}{4\pi}\textrm{Tr}( \partial_+ gg^\dagger \bm\tau),\quad \mathbf J_{R,s}=-\frac{i}{4\pi}\textrm{Tr}(g^\dagger  \partial_{-} g\bm\tau),
\ee
where $\partial_\pm=\partial/\partial x_\pm$. The theory for the low energy sector of the Hubbard model is   equivalent to that of the Heisenberg spin chain, the only distinction being in the spin velocity $v_s$, which depends on $U$. 

\subsection{Edge singularities at high energies: imposing spin $SU(2)$ invariance in spin correlation functions}

Although low energy theories based on the linear dispersion approximation yield reliable results for thermodynamic quantities, in general they fail to predict the correct  edge singularities of dynamic correlation functions.\cite{RMPimamb}  For this purpose it is important to take into account formally irrelevant perturbations that break the Lorentz invariance of  the  fixed point Hamiltonian. Nonlinear Luttinger liquid theory makes progress by refermionizing the elementary excitations.\cite{imambekovscience}  For spin-1/2 models, this means  defining spinless fermions associated with holon and spinon bands that have a finite  curvature about the Fermi points.\cite{sela,schmidt}

The idea behind the effective impurity models for edge singularities is the same for all dynamic correlation functions. Essentially, it involves defining high energy sub-bands within the dispersion of elementary excitations, in addition to the chiral low energy modes.\cite{pustilnik} The single-particle states used to define the high energy sub-bands depend on the momentum and energy of interest for the dynamic response function.  In order to motivate the application  of the $SU(2)$ invariant effective field theory for edge singularities, let us turn for the moment to the case of \emph{spin} correlation functions, for which more is known concerning the implications of   $SU(2)$ invariance.\cite{pereira08,imambsu2}  We will show that the proposed definition of a high energy impurity spinor in Eq. (\ref{Dsspinor}) below recovers known results. 

\subsubsection{Lower edge of the two-spinon continuum}

\begin{figure}
\begin{center}
\includegraphics*[width=80mm]{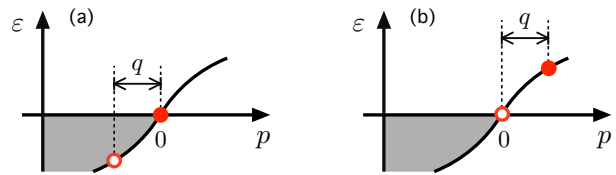}
\caption{(Color online.) (a) ``Deep hole'' particle-hole excitation that gives the lower edge of the longitudinal spin DSF. In the effective field theory, spinons are interacting  spinless fermions with   hole states for $-\pi<p<0$ and particle states for $0<p<\pi$. (b) Particle-hole excitation with high energy particle. Due to spin inversion symmetry, the dispersion is particle-hole symmetric and (a) and (b) are degenerate. \label{figspinband}}
\end{center}
\end{figure}

For the half-filled Hubbard model  with $U>0$, the spectrum of spin correlation functions is gapless. The effective  theory for edge singularities of the spin DSF has been worked out for the  XXZ model,\cite{pereira08,imambsu2}  which only has $U(1)$ symmetry for general anisotropy parameter but includes the $SU(2)$ symmetric Heisenberg point. In the spinless fermion language, the spin excitations are described by  particles and holes in an interacting band (see Fig. \ref{figspinband}). The longitudinal spin DSF   is defined as \be
S^{zz}_s(q,\omega)=\sum_je^{-iqj}\int_{-\infty}^\infty dt\,e^{i\omega t}\langle S_j^z(t)S_0^z(0)\rangle.\ee
We can also consider the transverse spin DSF\be
S^{+-}_s(q,\omega)=\sum_je^{-iqj}\int_{-\infty}^\infty dt\,e^{i\omega t}\langle S_j^+(t)S_0^-(0)\rangle.\ee
Spin $SU(2)$ invariance at zero magnetic field implies $S^{zz}_s(q,\omega)=S^{+-}_s(q,\omega)/2$.

The lower edge of the support of  $S_s^{zz}(q,\omega)$ corresponds to  the lower threshold of the two-spinon continuum and  is described as a ``deep hole'' excitation with a hole with momentum $p=-q$ below the Fermi point and a particle exactly at the Fermi point. The energy of this excitation is equal to the spinon mass shell $\varepsilon_s(q)>0$. Since at zero magnetic field the spin band is particle-hole symmetric,\cite{pereira08} the excitation with a hole at the Fermi point and a particle at $p=q$ above the Fermi point is degenerate with the deep hole excitation.

The edge singularity in this case is described by a $q$-dependent effective model which, besides the low energy states near the Fermi points, contains impurity sub-bands associated with the deep hole or the high energy particle.  The spin DSF  shows a singularity above the spinon mass shell, $S^{zz}_s(q,\omega)\sim\delta\omega^\mu$, with $\delta\omega= \omega-\varepsilon_s(q)$.  The lower edge exponent $\mu$ is determined by the scaling dimension of the operator that creates the particle-hole excitations after performing a  unitary transformation that decouples the impurity modes from the bosonized Fermi surface modes. For details, see Ref. \onlinecite{RMPimamb}. After this unitary transformation, up to irrelevant operators, the effective Hamiltonian density assumes the noninteracting form $\mathcal{H}=\mathcal{H}_s^{(0)}+\mathcal{H}_d^{(0)}$, where \bea
\mathcal{H}_s^{(0)}&=&\frac{v_s}2[(\partial_x\varphi_{R,s})^2+(\partial_x\varphi_{L,s})^2],\label{Hsdafter1}\\
\mathcal{H}_d^{(0)}&=&d_s^\dagger (\varepsilon_s-iu_s\partial_x)d_s+\bar d_s^\dagger (\varepsilon_s-iu_s\partial_x)\bar d_s.\label{Hsdafter2}
\eea 
Here,  $d_s(x)$ and $\bar d_s(x)$  are  field operators that annihilate a high energy spinon particle  and a deep spinon hole, respectively, and $u_s<v_s$ is the velocity of both impurity sub-bands. The high energy sub-bands are defined with momenta centred at $\pm q$ and have momentum cutoff $\Lambda$, with $u_s \Lambda\ll1$ (see Fig. \ref{figspinband}).The ground state is a vacuum of $d_s$ and $\bar d_s$. After the unitary transformation, the spin operator that is applied to the ground state is of the form\bea
S^z(x)&\propto&d_s^\dagger(x)e^{-i\sqrt{2\pi}\gamma_R\varphi_{R,s}(x)}e^{-i\sqrt{2\pi}\gamma_L\varphi_{L,s}(x)}\nonumber\\
&&- \bar d_s^\dagger(x)e^{i\sqrt{2\pi}\gamma_R\varphi_{R,s}(x)}e^{i\sqrt{2\pi}\gamma_L\varphi_{L,s}(x)},\eea
where the relative minus sign between the two terms comes from ordering the Klein factors of the sub-bands (recall $\bar d_s^\dagger$ creates a hole). For the $U(1)$ symmetric model, the parameters $\gamma_{R,L}$ can be related to exact phase shifts.\cite{pereira08} In the case of $SU(2)$ symmetry, these parameters can be fixed by the condition that  longitudinal and transverse spin correlations have the same exponents.\cite{imambsu2} This condition implies $\gamma_R=1/\sqrt2$ and $\gamma_L=0$ and the $z$ component of the spin operator reduces to \be
S^z(x)\propto d_s^\dagger(x)e^{-i\sqrt{\pi}\varphi_{R,s}(x)}- \bar d_s^\dagger(x)e^{i\sqrt{\pi}\varphi_{R,s}(x)}.\label{szwithgs}
\ee
The dimension-1/4 vertex operators in Eq. (\ref{szwithgs}) can be recognized as the components of the chiral spinor   $g_R$ in Eq. (\ref{galpha}). This observation motivates regarding $ d$ and $\bar d$ as the components of a high energy spin impurity spinor\be
D_s(x)= \left( {\begin{array}{c}
 d_s(x) \\
  \bar d_s(x)\\
 \end{array} } \right),\label{Dsspinor}
\ee
which must transform under spin $SU(2)$ rotations as $D_{s,i}\to D_{s,i}^\prime=U_{ij}D_{s,j}$. With this definition, the particle-hole degree of freedom of the impurity is interpreted as an   effective pseudospin 1/2. The operator in Eq. (\ref{szwithgs}) can be rewritten in the compact form\be
S^z(x) \propto D_s^\dagger(x)\tau^z g_R(x).
\ee
In fact, the equivalence of longitudinal and transverse correlation functions follows from the correlation functions of the spin $SU(2)$ vector operator\be
\mathbf{S}(x)\propto  D_s^\dagger(x)\bm \tau g_R(x).\label{spinvector}
\ee
The transverse components in Eq. (\ref{spinvector}) also agree with known results.\cite{imambsu2,sela,hamed}

The free Hamiltonian in Eqs. (\ref{Hsdafter1}) and (\ref{Hsdafter2}) can be rewritten in the $SU(2)$ invariant form\be
\mathcal{H}=\frac{2\pi v_s}3(\mathbf{J}_{R,s}^2+\mathbf{J}_{L,s}^2)+D^\dagger_s(\varepsilon_s-iu_s\partial_x)D^{\phantom\dagger}_s.\label{H0invariant}
\ee
In this effective model for the lower edge singularity, the states in the Hilbert space are constrained to have either zero (ground state) or one impurity (excited states), $N_{d,s}=\int dx\, D^\dagger_s(x)D^{\phantom\dagger}_s(x)=0,1$.   There is no essential distinction between the two high energy sub-bands since the transverse components of the total spin vector\be
\mathbf S=\int dx\,[\mathbf J_{R,s}(x)+\mathbf J_{L,s}(x)+D_s^\dagger (x)(\bm\tau/2) D_s^{\phantom\dagger}(x)] 
\ee
generates rotations of deep holes into   high energy particles. The time ordered propagator for the free $D_s$ field reads\be
 \langle T D^{\phantom\dagger}_{s,i} (x,t)D_{s,j}^\dagger(0,0)\rangle=\delta_{i,j}\theta(t)e^{-i\varepsilon_s t}\delta(x-u_st),\label{propagD}
\ee
where $D_{s,1}= d_s$ and $D_{s,2}= \bar  d_s$. The correlation functions for the chiral spinors are given by the standard conformal field theory result \bea
\langle g^\dagger_{L,i}(x,t)g^{\phantom\dagger}_{L,j}(0,0)\rangle&\propto &\delta_{i,j}(x_+)^{-1/2}\\
\langle g^\dagger_{R,i}(x,t)g^{\phantom\dagger}_{R,j}(0,0)\rangle&\propto& \delta_{i,j}(x_-)^{-1/2}.\label{gCFT}\eea  
Using these expressions, we can calculate the edge exponent $\mu$ from $S^{zz}_s(q,\omega)\sim \int dx\int dt\,e^{i\omega t}\langle B(x,t)B^\dagger(0,0) \rangle$ with $B^{z\dagger}(x)\propto D_s^\dagger(x)\tau^z g_R(x)$. This gives $\mu=-1/2$, the same as the result for the Heisenberg model.\cite{pereira08}

In order to connect with the methods developed for $U(1)$ symmetric models, Hamiltonian (\ref{Hsdafter1}) must be interpreted as the effective model \emph{after} the unitary transformation that decouples the mobile impurity. However, a different approach could be to write down Eq. (\ref{H0invariant}) directly based only on $SU(2)$ symmetry. In this case, in addition to the terms in Eq. (\ref{H0invariant}) we would be led to write down the marginal operator \be
\delta\mathcal{H}_{RLD}=-2\pi v_s(\kappa_R \mathbf J_{R,s}+\kappa_L \mathbf J_{L,s})\cdot D_s^\dagger\bm\tau D_s^{\phantom\dagger},\label{intRLD}
\ee
where $\kappa_{R,L}$ are dimensionless coupling constants. The longitudinal part of this operator amounts to a  density-density interaction between the impurity and the Fermi surface modes. The full operator $\delta\mathcal{H}_{RLD}$ is equivalent to a two-channel Kondo coupling, which appears naturally in the problem of a mobile spin-1/2 impurity coupled to a 1D electron gas.\cite{lamacraft} In Appendix \ref{App:RG} we show that the $\kappa_{R,L}$ operators are \emph{marginally irrelevant} for $\kappa_{R,L}>0$ (equivalent to ferromagnetic Kondo coupling). Although we are not able to derive the bare coupling constants starting from the Hubbard model for general $U$, we shall assume that $\kappa_{R,L}$ are positive for $U>0$ because otherwise we would not recover the known results for the Heisenberg model.  Moreover, it is known that the finite size spectrum for excited states of the  Hubbard model that contain high energy holes in the spin band fits the ``shifted'' conformal field theory form,\cite{essler10}  suggesting that the marginal operator should be irrelevant for any finite $U$. With the asymptotic decoupling of the impurity spinon, the symmetry of the effective  model (\ref{H0invariant}) becomes $SU(2)_L\times SU(2)_R\times SU(2)_D$.

\begin{figure}
\begin{center}
\includegraphics*[width=80mm]{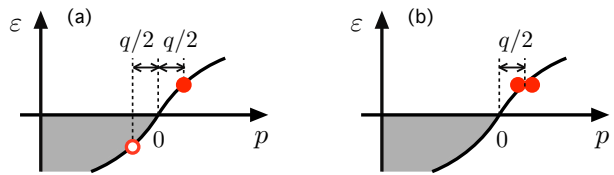}
\caption{(Color online.) (a)  Particle-hole excitation that gives the upper edge of the two-spinon continuum in the longitudinal spin DSF.  (b)  $S^z=+1$ excitation     corresponding to the creation of two spinons above the Fermi points. Due to spin inversion symmetry, the latter is degenerate with the $S^z=0 $ excitation in (a).
\label{figexciton}}
\end{center}
\end{figure}

\subsubsection{Upper edge of the two-spinon continuum}

The $SU(2)$ invariant effective theory  can also be applied to the  upper edge of the two-spinon continuum, where it is known that  the spin DSF for the Heisenberg model has another power law singularity.\cite{pereira08} In this case the threshold is given by a particle with momentum $q/2$ and a hole with momentum $-q/2$, as shown in Fig. \ref{figexciton}a. In this case the excited state has two impurities.  The particle and hole states form the components of  a single  impurity spinor $D_s$ as given by  Eq. (\ref{Dsspinor}). Thus $N_{d,s}=\int dx\, D^\dagger_s(x)D^{\phantom\dagger}_s(x)=2$ for excited states. We introduce the time reversal conjugated spinor \be
D^*_s(x)=\left( {\begin{array}{c}
 - \bar d^\dagger_s(x) \\
  d^\dagger_s(x)\\
 \end{array} } \right),\label{4spinor}
\ee
which transforms like $D_s$ under spin rotations. 
The excited state that describes the upper threshold of the two-spinon continuum   is created by acting on the ground state with the operator $B^{z\dagger}(x)\propto D^\dagger_s(x)\tau^z D^*_s(x)$, where the high energy particle and high energy hole in the final state must be treated as distinguishable particles, as in a two-body problem.\cite{pereira08}
$SU(2)$ symmetry dictates that the effective impurity model is of the form\bea
\mathcal{H}&=& D^\dagger_s\left(\varepsilon_s-iu_s\partial_x-\frac{\partial_x^2}{2m_s} \right) D^{\phantom\dagger}_s+V_{s} ( D^\dagger_s  D^{\phantom\dagger}_s)^2\nonumber\\
&&+\frac{2\pi v_\nu}3(\mathbf{J}_{R,s}^2+\mathbf{J}_{L,s}^2).\label{Hexciton}
\eea
Here we have included the parabolic term in the dispersion of the impurities, with effective mass $m_s<0$.

The marginal part of the $V_{s}$ operator in Eq. (\ref{Hexciton}) acts on the excited state as a density-density interaction between the two impurities. For $U>0$, we expect  $V_{s}<0$ as obtained for the Heisenberg model,\cite{pereira08} implying an attractive interaction between particle and hole.  The    $V_{s}$ interaction turns out to be crucial for the upper edge singularity  $S_s^{zz}(q,\omega)\sim \delta\omega^\mu$, with $\delta\omega=2\varepsilon_s(q/2)-\omega$. For $V_s=0$,  the density of states diverges as $\delta\omega\to0$ due to the Van Hove singularity for particle and hole with equal velocities. However, for any $V_s\neq0$, the solution of the two-body problem shows that  the matrix elements  are strongly affected by resonant scattering and turn the divergence into a square-root cusp with $\mu=+1/2$. The effect is analogous to a 1D exciton problem for particles with negative mass, hence no particle-hole bound state above the continuum for $V_s<0$.

But what we have described is the interpretation of the singularity in the longitudinal spin DSF. An alternative route to determine the edge exponent would be to rely on the spin $SU(2)$ symmetry and consider the transverse spin DSF. In this case, instead of a particle-hole pair, the excited state has either two particles with momentum $q/2$ (for $S_s^{-+}(q,\omega)$) or two holes with momentum $-q/2$ (for $S_s^{+-}(q,\omega)$) (see Fig. \ref{figexciton}b). The excited state with $S^z=+1$ is created by the operator $B^{+\dagger}(x)\propto D^\dagger_s(x+\frac\varepsilon2)\tau^+D^{*}_s(x-\frac\varepsilon2)\sim {d_s^\dagger \partial_x d_s^{\dagger}}$. In the case of the transverse components $B^{\pm\dagger}$, we need to introduce the point splitting  because the operator creates two spinless fermions with approximately the same momentum. Thus the leading term  has higher scaling dimension than the longitudinal component $B^{z\dagger}$. On the other hand, for spinless fermions the $V_{s}$ interaction is irrelevant --- the s-wave scattering amplitude vanishes ---  and can be neglected in the effective Hamiltonian. Remarkably, we encounter the same exponent $\mu=1/2$ due to matrix elements for free spinless fermions with vanishing relative momentum.\cite{penc00} This can be verified by calculating the propagator for the pairing field $ {d_s^\dagger \partial_x d_s^{\dagger}}$.\cite{pereira10}   Therefore, $SU(2)$ symmetry tells us that the upper edge exponent can be interpreted as  due to either strong interactions in the excitonic pair or statistics of free spinless fermions.

\subsection{Edge singularities at high energies: imposing $\eta$-spin $SU(2)$ invariance in the charge DSF at half filling}

We now turn to edge singularities in $S(q,\omega)$, which involve the creation of high energy holons. Within the field theory approach, we represent the charge excitations as holes in a completely filled band or particles in an empty band, with Mott-Hubbard  gap $2\Delta$. We will   borrow the nomenclature  often adopted in the literature and  refer to these  bands as the lower Hubbard band and the upper Hubbard band, respectively. Since there are no Fermi points in this case, the holon band only contributes with impurity sub-bands to the effective model.  By analogy with     $ D_s$   in Eq. (\ref{Dsspinor}), we define  the charge impurity spinor for given high energy holon sub-bands as\be
D_c(x)= \left( {\begin{array}{c}
 d_c(x) \\
  \bar d_c(x)
 \end{array} } \right),\label{Dcspinor}
\ee 
such that  $d_c^\dagger$ creates a   particle in the  upper Hubbard band and $\bar d_c^\dagger$ creates a hole in the  lower Hubbard band. The ground state is a vacuum of $D_c$. Due to $\eta$-spin $SU(2)$ symmetry, explicit in Eq. (\ref{Sqwinvariant}), the effective Hamiltonians as well as the operators that create high-energy excitations in the field theory must be written in terms of the charge impurity spinor. The generator of $\eta$-spin rotations is represented by\be
\bm\eta=\int dx\,D^\dagger_c(x)(\bm\tau/2) D^{\phantom\dagger}_c(x).
\ee

We are now in a position to compute the exponents for the thresholds  of the charge DSF in Fig. \ref{continuum49}. 

\subsubsection{Boundary line $\omega_{2c}^{-}(q)$ for $q<q_\diamond$}

\begin{figure}
\begin{center}
\includegraphics*[width=80mm]{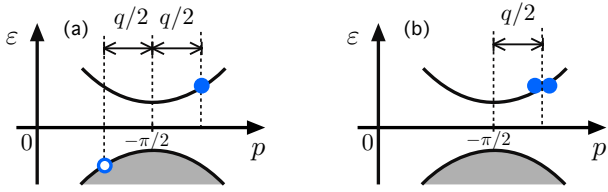}
\caption{(Color online.) (a) Particle-hole excitation that gives the lower edge of the two-holon continuum in the charge DSF for $q<q_\diamond$. In the effective field theory, holons are spinless fermions with a gap between  the lower Hubbard band and the upper Hubbard band. (b)  $\eta^z=+1$ excitation that adds two particles to the upper Hubbard band. Due to particle-hole symmetry, the latter is degenerate with the $\eta^z=0 $ excitation in (a).
\label{figholonband}}
\end{center}
\end{figure}

Consider first the lower edge of the two-holon continuum for momentum in the range $q<q_\diamond$, such that $\omega_{2c}^{-}(q)=2\varepsilon_c(-\pi/2+q/2)$. The effective model in this case has two charge impurities in the excited state, $N_{d,c}=\int dx\,D^\dagger_c(x)D^{\phantom\dagger}_c(x)=2$. The $\eta^z=0$ state corresponds to a hole in the lower Hubbard band and a particle in the upper Hubbard band,  as illustrated in Fig. \ref{figholonband}a. The particle and hole are the components of the same $D_c$ spinor and the situation is analogous to the upper edge of the two-spinon continuum in the spin DSF. Due to the $\eta$-spin $SU(2)$ symmetry in Eq. (\ref{Sqwinvariant}), the edge exponent can also be calculated from the excited state of two  particles created in the same sub-band (Fig. \ref{figholonband}b). The vector operator that creates these $\eta$-spin triplet excitations is $\mathbf B^\dagger(x)\propto D^\dagger_c(x+\frac\varepsilon2)\bm\tau  D^{*}_c(x-\frac\varepsilon2)$. The effective Hamiltonian density consistent with $SU(2)\times SU(2)$ symmetry reads\be
\mathcal H= D^\dagger_c\left(\varepsilon_c-iu_c\partial_x-\frac{\partial_x^2}{2m_c} \right) D^{\phantom\dagger}_c+V_{c}(D^\dagger_c D^{\phantom\dagger}_c)^2.\label{H2clower}
\ee
Due to symmetry, there is no coupling between holons and low-energy spinons  at the level of marginal operators. 
Since $m_c>0$, we expect $V_c>0$ for absence of a particle-hole bound state below the threshold. It follows that the edge singularity is of the form $S(q,\omega)\sim \delta\omega^\mu$ with $\delta\omega=\omega-\omega_{2c}^-(q)$ and $\mu=1/2$. A similar conclusion can be reached for the singularity at the upper edge of the two-holon continuum $\omega_{2c}^+(q)$ for all values of $q$. We note  that   $\eta$-spin rotations mix states with $\eta^z=0,\pm1$, but the total momentum of the $\eta^z=0$ state differs from the momentum of the $\eta^z=\pm1$ states by $\pi$. This is consistent with the spectrum from the BA.\cite{esslerbook} 

\subsubsection{Boundary line $\omega_{2c2s}^{-}(q)$ for $q_\triangleleft<q<q_\triangleright $}

For $q_\triangleleft<q<q_\triangleright=\pi-q_\triangleleft$, the lower edge of the support  of $S(q,\omega)$ has one low energy spinon and one impurity spinon  in addition to the two holons. The operator that creates this two-holon-two-spinon excitation must be constructed using one low energy  chiral spinor, one  $D_s$ spinor and two $D_c$ spinors. Furthermore, selection rules impose that the operator is a vector of $\eta$-spin rotation and a scalar of spin rotation. These conditions naturally lead to   $\mathbf B^\dagger(x)\propto D^\dagger_c(x+\frac\varepsilon2) \bm \tau D^{*}_c(x-\frac\varepsilon2)D^\dagger_s(x)g_R(x) $ as the operator with the lowest scaling dimension. Besides the sum of Eqs. (\ref{H0invariant}) and   (\ref{H2clower}) with $u_c=u_s=u$, the effective  Hamiltonian contains the symmetry allowed interaction between the spin impurity and the charge impurities \be
\delta\mathcal H_{cs}=V_{cs}D_c^\dagger D_c^{\phantom\dagger}D_s^\dagger D_s^{\phantom\dagger}.
\ee
The parameter $V_{cs}$ could in principle be related to the exact phase shift in the nontrivial  $S$ matrix between a high energy holon and a high energy spinon. We then need to compute the propagator for three impurities that move with the same velocity, interact among themselves    but are decoupled from the low energy modes. It is easiest to discuss the $\eta^z=+1$ excitation instead of the $\eta^z=0$ one, trading the interactions between distinguishable charge hole and charge particle by the problem of noninteracting holons which are indistinguishable fermions. Simple power counting in the correlation function for $\mathbf B^\dagger(x)$ (the calculation is detailed in appendix \ref{app:exp32}) yields the edge singularity $S(q,\omega)\sim \delta \omega^\mu$ with $\delta\omega=\omega -\omega_{2c2s}^-(q)$ and $\mu=3/2$.

\subsubsection{Boundary line $\omega_{2c2s}^{-}(q)$ for $q_\triangleright<q<\pi$}
 
For $q_\triangleright<q<\pi$, the lower edge of the support has two  spinons at opposite Fermi points. Thus we are looking for a spin scalar operator that involves the low energy modes only. The  momentum $\pi$ scalar operator of the WZW model is the trace of the matrix field Tr$[g(x)]$, which has scaling dimension $1/2$. The operator that creates the excitation  in this case is then $\mathbf B^\dagger(x)\propto  D^\dagger_c(x+\frac\varepsilon2) \bm \tau D^{\phantom\dagger}_c(x-\frac\varepsilon2)$Tr$[g(x)]$. Again, we find   the edge exponent $\mu=3/2$.


\subsubsection{Boundary line $\omega_{2c}^{-}(q)$ for $q_\diamond<q<\pi$}

Finally, let us discuss the lower edge of the two-holon continuum for $q_\diamond<q<\pi$. In this case the $\eta^z=0$ excited state has a hole in lower Hubbard band and a particle in the upper Hubbard band that move with the same velocity, but are \emph{not} associated with the same charge impurity  spinor. We denote the  spinor for the holon with momentum below the inflection point of the holon dispersion (see Fig. \ref{fig:holonspinon}) by $D_c$ and the spinor for the   holon above the inflection point  by $\tilde D_c$. The occupation of the impurity sub-bands in the excited state is $N_{d,c}=\int dx\,D^\dagger_c(x)    D^{\phantom\dagger}_c(x)=1 $ and $\tilde N_{d,c}=\int dx\,\tilde D^\dagger_c(x)  \tilde D^{\phantom\dagger}_c(x)=1 $. The vector operator in this case reads $\mathbf B^\dagger(x)\propto D_c^\dagger(x)\bm\tau  \tilde D^*_c(x)$. The effective Hamiltonian density with marginal operators allowed  by symmetry is\bea
\mc{H}& =& D^\dagger_c\left(\varepsilon_c-iu_c\partial_x-\frac{\partial_x^2}{2m_c} \right) D^{\phantom\dagger}_c\nonumber\\
&&+\tilde D^\dagger_c\left(\tilde\varepsilon_c-iu_c\partial_x-\frac{\partial_x^2}{2\tilde m_c} \right)\tilde D^{\phantom\dagger}_c\label{CandE}\\
&&+V_{c}^{C}D^\dagger_c   D^{\phantom\dagger}_c\tilde D^\dagger_c   \tilde D^{\phantom\dagger}_c +V_{c}^{E}D^\dagger_c  \bm\tau D^{\phantom\dagger}_c\cdot \tilde D^\dagger_c \bm \tau  \tilde D^{\phantom\dagger}_c ,\nonumber
\eea
where $V_c^C$  and $V_c^E$ are the Coulomb and exchange interactions between the distinguishable impurities. (For models (\ref{Hexciton}) and (\ref{H2clower}) with a single impurity spinor, these two interactions are equivalent.)

The model in Eq. (\ref{CandE}) is again similar to a 1D exciton problem. There is a Van Hove singularity in the density of states when the relative momentum between $D_c$ and $\tilde D_c$ holons approaches zero. We expect this divergence to be removed for arbitrarily weak final-state interactions.   There is a priori no reason why $V_c^{C}$ and $V_c^E$ should be zero or even small at finite $U$. Depending on the sign of the effective scattering amplitude, a bound state can be formed below the continuum, which is in fact observed numerically for the extended Hubbard model.\cite{jeckelmann} 

However, in appendix \ref{integrable} we show that the existence of nontrivial conservation laws in the Hubbard model requires   $V_c^C=V_c^E=0$ exactly.  Remarkably, the integrability of the model implies that impurity holons associated with different $D_c$ spinors do not scatter off each other. 

We stress that   the vanishing of $V_c^C$ and $V_c^E$ does not follow from $\eta$-spin $SU(2)$ symmetry alone. This is reasonable  because it is possible to generate infinitely many       models with the same symmetry that are not integrable, for instance by adding  finite range $\eta$-spin exchange interactions $\sum_{j,j^\prime} J_{j,j^\prime}\bm\eta_j\cdot  \bm\eta_{j^\prime}$ to the Hubbard model. For  non-integrable models, we generically expect   the formation of two-holon bound states\cite{penc00} below $\omega_{2c}^-(q>q_\diamond)$ --- as well as the broadening of any power-law singularity that is not protected by kinematics.

\begin{table}[b]

\caption{Predictions of the $SU(2)$ invariant effective impurity models for the charge DSF of the Hubbard model. The boundary lines considered here are illustrated in Fig. \ref{continuum49}. As the frequency approaches the boundary lines, $\delta\omega\to0$, the  DSF behaves like $S(q,\omega)\propto \int dx\int dt\, e^{i\omega t}\langle \mathbf B(x,t)\cdot \mathbf B^\dagger (0,0)\rangle\sim \delta\omega^\mu$.  \label{tableII}}

\begin{tabular}{c|c|c}
\hline\hline 
Boundary line&Vector operator $\mathbf  B^\dagger$&Edge exponent $\mu$
 \\
\hline 
$\omega_{2c}^-(q<q_\diamond)$&$D^\dagger_c\bm\tau D_c^{*}$&$1/2$\\
$\omega_{2c}^-(q>q_\diamond)$&$D^\dagger_c\bm\tau \tilde D_c^{*}$&$-1/2$\\
$\omega_{2c}^+(q)$&$D^\dagger_c\bm\tau D_c^{*}$&$1/2$\\
$\omega_{2c2s}^-(q_\triangleleft <q<q_\triangleright)$&$D^\dagger_c\bm\tau D_c^{*}D^\dagger_s g_R$&$3/2$\\
$\omega_{2c2s}^-(q>q_\triangleright)$&$D^\dagger_c\bm\tau D_c^{*}$Tr$[g]$&$3/2$\\
\hline\hline 

\end{tabular}
\end{table}

When we set $V_c^C=V_c^E=0$, the propagator of $\mathbf B^\dagger(x)\propto  D^\dagger_c(x)\bm\tau \tilde{ D}^*_c(x)$ factorizes into free propagators for $D_c$ and $\tilde D_c$ impurities. The Van Hove singularity of the density of states persists in the DSF as $S(q,\omega)\sim \delta\omega^\mu$ with $\delta\omega=\omega-\varepsilon_c-\tilde\varepsilon_c$ and $\mu=-1/2$. This is the only divergent edge singularity in the charge DSF and only appears at \emph{finite energies and finite} $U$.

The results for the boundary lines discussed in this section are summarized in Table \ref{tableII}. The exponents for the lines $\omega_{2c}^-(q<q_\diamond)$ and $\omega_{2c}^+(q)$ are consistent  with the large-$U$ results of Ref. \onlinecite{penc00}. The exponent for the line $\omega_{2c2s}^-(q_\triangleright<q<\pi)$ agrees with the low energy result obtained assuming $v_c=v_s$ in Ref. \onlinecite{controzzi}. Our results show that these exponents hold at finite $U$ and away from the low energy limit. The exponents for the lines $\omega_{2c}^-(q_\diamond<q<\pi)$ and $\omega_{2c2s}^-(q_\triangleleft<q<q_\triangleright)$ could not be obtained by either large-$U$ or low energy approximations. Notice that the exponents predicted by the $SU(2)$ invariant impurity models are all half-integers, in contrast with the continuously varying exponents of the metallic phase.\cite{carmelo,essler10}

\section{Numerical results \label{sec:DMRG}}

\subsection{Methods}

We have used the tDMRG method to compute the real time density-density
correlation function $G(j,t)=\langle n_j(t) n_0(0)\rangle$ for Hubbard chains
with open boundary conditions and  lengths up to 200 sites.  The method starts
with a traditional DMRG calculation,\cite{White92,White93} obtaining the ground state
$|GS\rangle$ of the finite chain.
The single site operator
$\eta^z_0$ for a central site 0 is applied to the ground state, and then this
state is evolved in real time, obtaining $|\psi(t)\rangle$.  The original ground state
is retained in matrix product state (MPS) form, so that the tDMRG need only target $|\psi(t)\rangle$.
At each time step we measure 
$\langle \eta^z_j(t) \eta^z_0(0)\rangle$ by measuring the off-diagonal MPS overlaps
$\langle GS | \eta^z_j | \psi(t) \rangle$ for all sites $j$.  A single run provides results for
all frequency and momenta by Fourier transforming over time and space (i.e. $j$).

The time evolution operator is written as a product of exact nearest-neighbor bond exponentials,
as in a familiar Suzuki-Trotter breakup. Recently Kirino, Fujii, and Ueda have reported excellent performance
with a particular fourth order breakup, in which every bond operator is applied in every half-sweep,
but in reverse order for every other half-sweep.\cite{ueda}
We have also found that this method gives very small finite-time-step error and appears
to be superior to other breakups for high accuracy calculations.

The main limitation of the tDMRG method is on the maximum time reached by the simulation, due
to the  growth of entanglement with running time. Typically, we have reached
$t_{\textrm{max}}\sim 20$ in units of inverse hopping, keeping a maximum of $m=2500-4000$ states. 
We find that the entanglement grows more rapidly for smaller values of
$U$ and this prevents us from studying $U<1$.  The spatial Fourier transform is done first,
and no windowing is required since within the maximum time reached, the signal which is
propagating within $|\psi(t) \rangle$ has not yet hit the edges of the system. 
Thus the resolution in momentum is not limited by the system size.
Windowing is
necessary in the time Fourier transform, but the frequency
resolution would be poor if we fit the window within $t_{\textrm{max}}$. Instead, we extrapolate
the time signal using linear prediction, allowing the use of a larger window.\cite{White08}
The resulting line shapes for the charge DSF do not have
any analytic input. 
A conservative estimate for the frequency
resolution of these line shapes is given by  $1/t_{\textrm{max}}$. 
This resolution could be substantially improved by using analytic results for the edge singularities of the 
DSF  to help extrapolate the DMRG data to much longer times.  

\subsection{tDMRG results for $S(q,\omega)$}

\begin{figure}
\begin{center}
\includegraphics*[width=75mm]{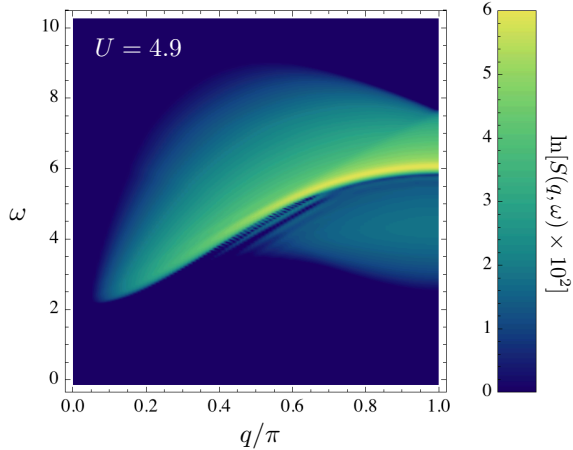}
\caption{(Color online.) Charge DSF of the Hubbard model at half filling  calculated by tDMRG
as function of momentum $q$ and energy $\omega$ for $U=4.9$. \label{fig:DMRGdensityplot}} \end{center}
\end{figure}

We now analyze   tDMRG results for $U=1$, $U=2$, and $U=4.9$, obtained without any analytic input, by comparing with the predictions of the field theory in Sec. \ref{sec:edgetheory} combined with the exact spectrum from the BA.

First we discuss  the result  for $S(q,\omega)$ for $U=4.9$   
shown in Fig. \ref{fig:DMRGdensityplot}. The exact support of the DSF 
in this case is illustrated in Fig. \ref{continuum49}; notice, however, that the energies in Fig. \ref{continuum49} are shifted by the Mott-Hubbard $2\Delta$ while the energies in Fig. \ref{fig:DMRGdensityplot} are not. The spectral weight
distribution in Fig. \ref{fig:DMRGdensityplot}  is consistent with the strong
coupling picture\cite{penc00} in the sense that the spectral weight is rather
small below the lower threshold of the two-holon continuum. However, for
values of $q$ near the zone boundary it is already visible that the onset of
the spectral weight occurs below the lower edge of the two-holon continuum. As
discussed in Sec. \ref{sec:model},  the main contribution to this weight is
due to excitations with two spinons in addition to two holons and the support
of $S(q,\omega)$ extends down to the line $\omega_{2c2s}^-(q)$.

\begin{figure}
\begin{center}
\includegraphics*[width=70mm]{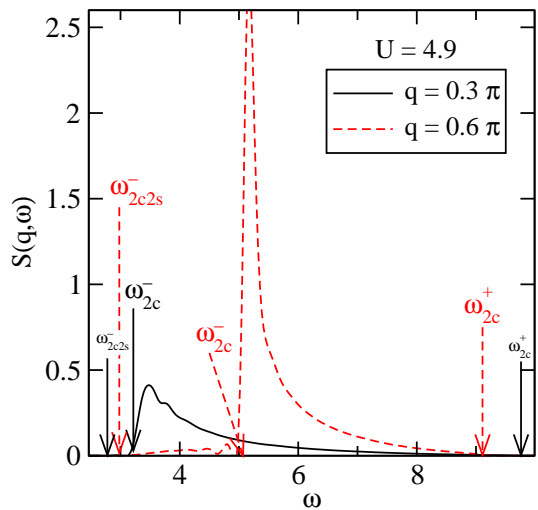}
\caption{(Color online.) Line shapes of $S(q,\omega)$ calculated by tDMRG for $U=4.9$ and two values of $q$. The arrows indicate the exact edges of the spectrum predicted by BA. The field theory predicts a square-root cusp at    $\omega=\omega_{2c}^-$ for $q=0.3\pi$ but a square-root divergence at $\omega=\omega_{2c}^-$ for $q=0.6\pi$. \label{fig:kU49twoqs}}
\end{center}
\end{figure}

Another featured observed in the tDMRG results for $U=4.9$ is a sharp
asymmetric peak above the lower edge of the two-holon continuum for $q$ near
the zone boundary. This effect is predicted by the theory in Sec.
\ref{sec:edgetheory} as a change in the exponent of the edge singularity from
$\mu=1/2$ for $\omega_{2c}^-(q<q_\diamond)$ to $\mu=-1/2$ for
$\omega_{2c}^-(q>q_\diamond)$. Using the exact holon dispersion for $U=4.9$, we
obtain $q_\diamond\approx 0.44\pi$. Fig. \ref{fig:kU49twoqs} shows constant-$q$ cuts of $S(q,\omega)$ for $q=0.3\pi<q_\diamond$ and $q=0.6\pi>q_\diamond$. The arrows indicate the threshold energies predicted by the BA.

In order to confirm the existence of two regimes for the $\omega_{2c}^-(q)$
edge exponent, we have analyzed the time decay of  the momentum dependent
correlation function $G(q,t)=\sum_j e^{-iqj}G(j,t)$.   We assume an asymptotic
power-law decay of $G(q,t)$ and fit the real part $G(q,t)$ in the  time range
$7<t<20$ to the formula
\be\label{fitform} \textrm{Re }G(q,t)=A_q
\cos(W_qt+\phi_q)t^{-\eta_q},
\ee
with $A_q,W_q,\phi_q,\eta_q$ as free parameters. Since $S(q,\omega)$ is given
by a time-frequency Fourier transform of $G(q,t)$, the exponent $\eta_q$ in
$G(q,t)$ is related to the exponent $\mu$ in $S(q,\omega)\sim \delta\omega^\mu$
by $\eta_q=1+\mu$ for  the smallest $\mu$ among the boundary lines.  The
fitting to Eq. (\ref{fitform}) should work best in the range $q_\diamond <q<\pi
$, in which we predict a square-root divergence in $S(q,\omega)$ which strongly
dominates the long time behavior of $G(q,t)$.

\begin{figure}
\begin{center}
\includegraphics*[width=75mm]{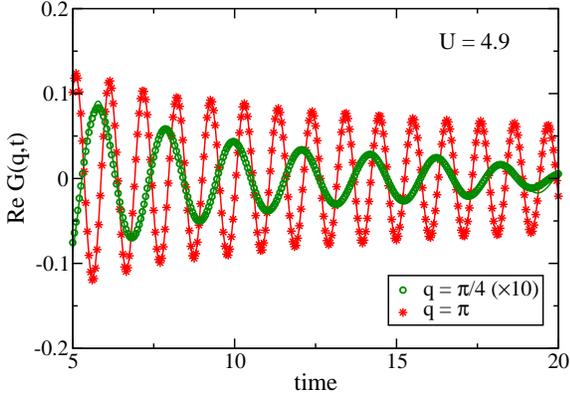}
\caption{(Color online.) Real part of the  density-density correlation function $G(q,t)=\sum_je^{-iqj}\langle n_j(t)n_0(0)\rangle$ for $U=4.9$. Symbols represent tDMRG data and solid lines are fits to the power-law decay form in Eq. (\ref{fitform}) for $7<t<20$. The data for $q=\pi/4$ has been rescaled by a factor of 10.  \label{figGqt}}
\end{center}
\end{figure}

\begin{figure}
\begin{center}
\includegraphics*[width=75mm]{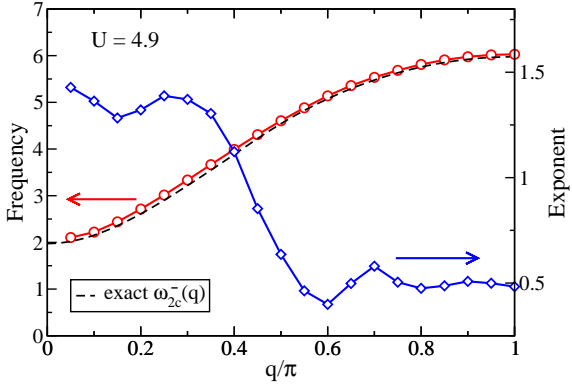}
\caption{(Color online.) Frequencies (circles) and exponents (diamonds)  obtained by fitting the tDMRG results for $G(q,t)$   for $U=4.9$   to Eq. (\ref{fitform}).  The dashed line represents the exact lower edge of the two-holon continuum.\label{fig:fitting}}
\end{center}
\end{figure}

The time decay of $\textrm{Re }G(q,t)$ is illustrated in Fig. \ref{figGqt}. The
energies and exponents obtained by  fitting the numerical results to Eq.
(\ref{fitform}) are shown in Fig. \ref{fig:fitting}. We first note that the
frequencies extracted from the tDMRG data are in excellent agreement with the
exact result from the BA. In fact, the tDMRG are slightly shifted to higher
energies as expected from the error due to the finite Trotter step.

Furthermore, the results for the exponent in Fig. \ref{fig:fitting} clearly
show  $\eta_q\approx 1/2$ for $q$ near the zone boundary. This supports the
existence of a square-root divergence in $S(q,\omega)$ which corresponds to  the
Van Hove singularity predicted by the theory in section \ref{sec:edgetheory} as  due to the absence of scattering between distinguishable impurities in the integrable model. We note that the existence of
a bound-state below the continuum would lead to a non-decaying contribution to
$G(q,t)$, which is not observed. On the other hand, the error in the numerical value of the exponent
increases with decreasing $q$, as the energy window of validity of the
square-root divergence in $S(q,\omega)$ decreases, which implies that longer
times would  be needed  in order to observe the asymptotic behavior of
$G(q,t)$. Nonetheless, Fig. \ref{fig:fitting} suggests that the exponent is
significantly larger  below $q=q_\diamond\approx 0.44\pi$. Recall that the
prediction of the effective impurity model is $\mu=1/2$ for $q<q_\diamond$,
which gives $\eta_q=3/2$.

\begin{figure}
\begin{center}
\includegraphics*[width=75mm]{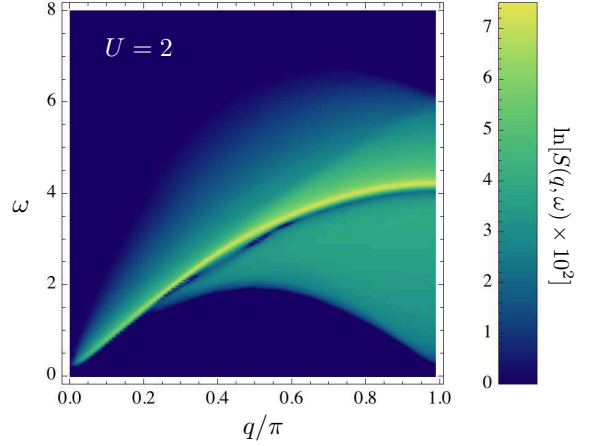}
\caption{(Color online.) Charge DSF  for $U=2$. \label{fig:DMRGdensityplot2}} \end{center}
\end{figure}

Let us now discuss the result for   $U=2$ shown in Fig. \ref{fig:DMRGdensityplot2}. For this smaller value of $U$, we see that a larger fraction of the spectral weight is located below the two-holon continuum. The lower edge of the support agrees with the exact line $\omega_{2c2s}^-(q)$ for $U=2$. We can quantify the distribution of spectral weight by computing an average frequency $\overline{\omega}(q)$ from the  first-moment sum rule as
\be\label{w1stm}
\overline\omega(q)=\frac{\int_0^\infty d\omega\, \omega S(q,\omega)}{\int_0^\infty d\omega\,  S(q,\omega)}=\frac{i\partial_tG(q,t=0)}{G(q,t=0)}.
\ee
The expression on the right hand side of Eq. (\ref{w1stm}) is directly provided
by  the tDMRG from the short time behavior of $G(q,t)$. For $U=4.9$ the average
frequency $\overline \omega(q)$ is always above $\omega_{2c}^-(q)$. In
contrast, for $U=2$ we find that $\overline \omega(q)<\omega_{2c}^-(q)$ for
$q\gtrsim 0.64\pi$. The difference $\omega_{2c}^-(q)-\overline \omega(q)$
increases as $q\to\pi$. Therefore, it appears that the   small $U$ behavior,
characterized by all the spectral weight lying below the two-holon continuum,
is approached more rapidly for larger values of $q$. 

\begin{figure}
\begin{center}
\includegraphics*[width=75mm]{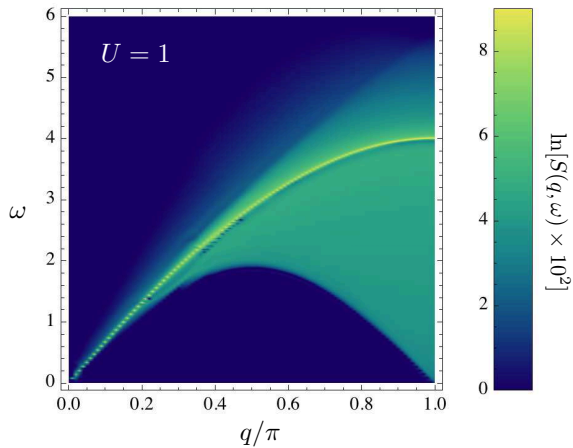}
\caption{(Color online.) Charge DSF  for $U=1$. \label{fig:DMRGdensityplot3}} \end{center}
\end{figure}

The transfer of spectral weight to below the two-holon continuum as $U$ decreases  is confirmed by the result for $U=1$ shown in Fig. \ref{fig:DMRGdensityplot3}. In this case the lines $\omega_{2c2s}^-(q)$ and $\omega_{2c}^-(q)$ are already very close to the lower and upper thresholds of the electron-hole continuum for $U=0$, respectively. However, there is still significant spectral weight in the two-holon continuum. 

\begin{figure}
\begin{center}
\includegraphics*[width=75mm]{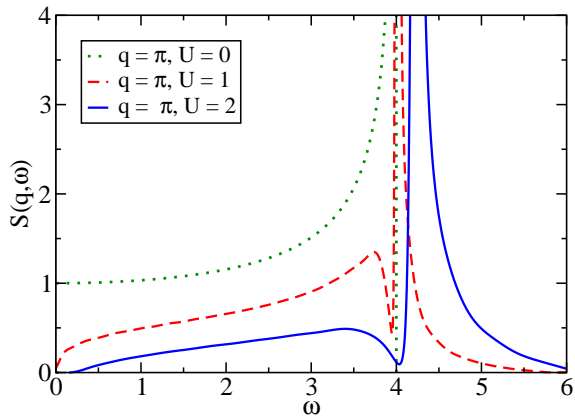}
\caption{(Color online.) Line shape of $S(q,\omega)$ calculated by tDMRG for $q=\pi$ and $U=1,2$. The exact free electron result for $U=0$ is also shown for comparison. \label{figsqw}}
\end{center}
\end{figure}

The results in Figs. \ref{fig:DMRGdensityplot2} and \ref{fig:DMRGdensityplot3} reveal that  $S(q,\omega)$
has a rounded peak below the lower edge of the two-holon continuum. The peak is
more clearly seen in Fig. \ref{figsqw}, which shows the line shape for $q=\pi$ for $U=1$ and $U=2$.
 Particularly in the case $U=1$ the peak is very narrow and the spectral weight is rapidly suppressed below  the onset of the two-holon contribution. We can also see that, although a large fraction of the total spectral weight is associated with
two-holon-two-spinon states, the singularity above $\omega_{2c}^-(q)$ persists.

\subsection{General picture for the DSF at finite $U$}

In light of the analytic results in Sec. \ref{sec:edgetheory}, the line shapes in Fig. \ref{figsqw} suggest a scenario for the $U$ dependence of $S(q,\omega)$. When combined with the exact spectrum from the BA, the $SU(2)$ invariant effective field theory  does not predict any divergence below the lower edge of the two-holon continuum. However, the free electron result in Eq. (\ref{freefermion}) exhibits a Van Hove singularity from \emph{below} the upper threshold of the electron-hole continuum. We interpret Fig. \ref{figsqw} as   indication that the free electron line shape is recovered  as  the  peak below $\omega_{2c}^-(q)$, which is rounded for any finite $U$, becomes narrower as $U\to0$. Only at $U=0$ does the Van Hove singularity develop at what is then the upper threshold of the electron-hole continuum.

Moreover, for any finite $U$ and fixed $q>q_\diamond$ (recall that $q_\diamond\to 0$ for $U\to0$) the square-root divergence above $\omega_{2c}^-(q)$ is always present. However, the spectral weight in the two-holon continuum vanishes for $U\to0$. The total spectral weight of $S(q,\omega)$ is not conserved as $U$ varies (see Eq. (\ref{sumrule})), but in relative terms the weight is transferred from the two-holon continuum for $U\to\infty$ to the subset of the  two-holon-two-spinon continuum that lies below the lower edge of the two-holon continuum for $U\to0$. 

The subset of the two-holon-two-spinon continuum that dominates $S(q,\omega)$ and reconstructs the electron-hole continuum in the limit   $U\to0$ can be obtained from the heuristic rule that  the holons are constrained to the minimum of the holon band (momentum $p_c=-\pi/2$ in  Fig. \ref{fig:holonspinon}), where the Mott-Hubbard gap closes for $U=0$, while the spinons are free to move along the spinon band. We conjecture that for $U\to0$ the matrix elements for the charge density operator  in Eq. (\ref{eq:Sqw}), which are not known except for small chains, select excited states with two holons and two spinons according to this rule. 

\begin{figure}
\begin{center}
\includegraphics*[width=85mm]{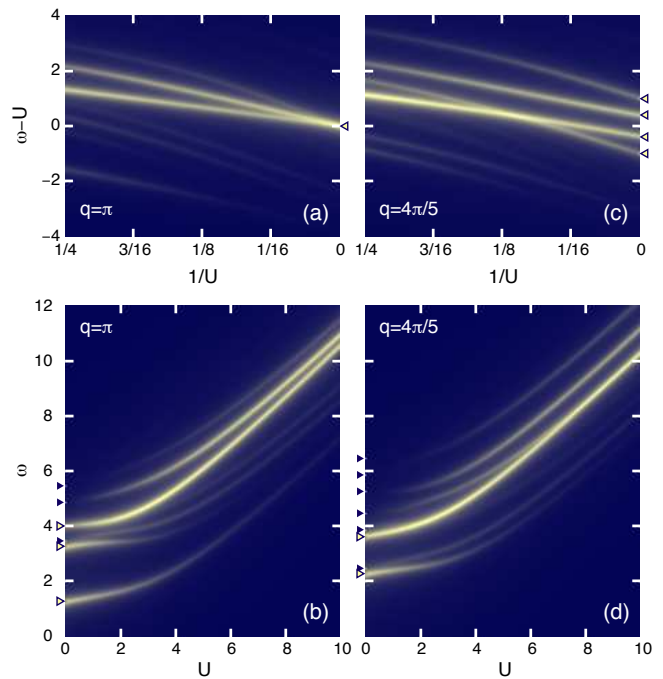}
\caption{(Color online.) Spectral weight for states that contribute to $S(q,\omega)$ for   a 10-site Hubbard ring for $q=\pi$ [(a) and (b)] and $q=4\pi/5$ [(c)  and (d)]. The center of each  line  represents the energy of an excited state $|\nu\rangle$ as a function of $1/U$ or $U$. The intensity is proportional to the matrix element $|\langle GS|n_q|\nu\rangle|$. In (a) and (c), the open triangles on the left-hand side mark the energies corresponding to   free spinless fermions in the large $U$ limit. In (b) and (d), the open triangles mark the energies of  one-particle-hole excitations in the $U=0$ case, and the filled triangles the energies of the two-particle-hole excitations.
 \label{figlanczos}}
\end{center}
\end{figure}

\subsection{Lanczos results for small systems}

In order to provide further evidence for the above scenario, we have calculated $S(q,\omega)$ for a 10-site half-filled chain with periodic boundary conditions  by exact diagonalization based on the  Lanczos method.  Figs. \ref{figlanczos}a and  \ref{figlanczos}b illustrates the    energies and matrix element for all  eigenstates of the Hamiltonian with total momentum $q=\pi$. The important point is that for this small system there is only one state that gives a large contribution to $S(q=\pi,\omega)$ in both limits of large $U$ and small $U$. This is the state that has energy equal to 4 at $U=0$, which corresponds to the maximum energy for an electron-hole excitation with $q=\pi$. 

By solving the Lieb-Wu equations\cite{esslerbook} for system size $L=10$, we have computed the exact energies of two-holon states  and identified that the state that evolves into the upper edge of the electron-hole continuum at $U=0$ is the lowest energy two-holon excitation.\cite{foot1} All   states with energy lower than the latter involve excitations in the spinon band. This observation is consistent with the  proposed  scenario for the $U$ dependence of $S(q,\omega)$ since it shows that the state  that defines the lower edge of the two-holon continuum and carries a large spectral weight splits off from the continuum below it for arbitrarily small $U$. In the thermodynamic limit we expect that this behavior corresponds to the disappearing of the Van Hove singularity below the upper edge of the electron-hole continuum and the formation of another Van Hove singularity above the lower edge of the two-holon continuum once we turn on the interaction.

We have also calculated the matrix elements for excitations with momentum $q=4\pi/5$ for the chain with $L=10$ (Figs. \ref{figlanczos}c and \ref{figlanczos}d).  Interestingly,  for $0<q<\pi$ there is a level crossing as a function of $U$ where the spectral weight associated with  the lowest energy two-holon state changes abruptly. This is a manifestation in the small system of the change in the nature of the lower edge of the two-holon continuum from $\mu=1/2$ to $\mu=-1/2$. The value of $U$ where the level crossing  happens is given by the condition $q_\diamond(U)=q$ at fixed $q$, where $q_\diamond(U)$ is twice the value of the momentum at the inflection point of the single holon dispersion. Indeed, in Fig. \ref{figlanczos}c we see that the weight in the lowest energy two-holon state is larger on the small $U$ side of the  level crossing ($1/U\gtrsim1/8$), which corresponds to the regime where we expect a square-root divergence above $\omega_{2c}^-(q)$ in the thermodynamic limit.

\section{Conclusion\label{sec:disc}}

In summary, we have studied the charge dynamic structure factor $S(q,\omega)$ of the Mott insulating phase of the 1D Hubbard model at finite $U$, based on a combination of Bethe ansatz, field theory and tDMRG. 

We used the BA solution to discuss the exact spectrum of   excitations that contribute to $S(q,\omega)$, without   low energy or strong coupling approximations. Unlike the metallic phase, the lower edge of the support of $S(q,\omega)$ is not given by the spinon mass shell, but by either the   lower edge of the two-holon continuum or the lower edge of the two-holon-two-spinon continuum that has three particles (two holons and one spinon) at finite energies with the same velocity.  In addition, an important difference  from the strong coupling theory is that at finite $U$ there is a range of momentum $q$ in which the lower edge of the two-holon continuum is described by two holons with the same velocity but different momenta. 

In order to investigate the behavior of the spectral weight of $S(q,\omega)$ near the edges of the spectrum, we relied on effective quantum impurity models. We have explicitly incorporated the $SO(4)$ symmetry of the Hubbard model at half filling by introducing $SU(2)$ spinors for the high energy charge and spin modes. The internal degree of freedom in these spinors stems from degenerate particle and hole sub-bands. Once we have these objects, we write down effective Hamiltonians with marginal operators that are allowed by the spin and $\eta$-spin $SU(2)$ symmetries.  In the effective impurity models the charge impurities are always decoupled from the low energy spin excitations due to   symmetry. On the other hand, the spin impurities are coupled to the low energy spin excitations, but the coupling is marginally irrelevant due to Kondo-type physics. 

The operators that are associated with each threshold are also identified using symmetry. These operators must have the lowest scaling dimension that is allowed by the conditions that the excited state has the correct number of impurities and that the operator has the correct quantum numbers for spin and $\eta$-spin rotations. In the case of $S(q,\omega)$, the operators are vectors of $\eta$-spin and scalars of spin rotations.  Due to the decoupling between low energy and high energy modes, the problem of edge singularities reduces to computing  few-body propagators for the high energy part, which can be affected by  final state interactions,  and combining them with  the  correlation functions for the low energy part, which are known from conformal field theory. Simple power counting in the time decay of the total correlation function then determines the edge exponent  $\mu$ for a given threshold. We have focused on $S(q,\omega)$, but the method can be readily applied to other dynamic response functions, such as the one-electron spectral function and the dynamic spin structure factor. 

The results of the effective quantum impurity models extend the validity of the low energy exponents\cite{controzzi} $\mu=1/2$ for $q\approx 0$ and $\mu=3/2$ for $q\approx\pi$   to the regime of finite $U$, even though the spectrum is not relativistic as in the sine-Gordon model. The impurity models combined with the exact spectrum from the BA also provide the range of $q$ over which these exponents hold. Remarkably, we found that the exponent $\mu=1/2$ at the lower edge of the two-holon continuum is verified only for $q<q_\diamond(U)$, where $q_\diamond(U)$ is determined by the inflection point of the holon dispersion relation. For $q>q_\diamond(U)$,  there is a Van-Hove type square-root divergence along the lower edge of the two-holon continuum, due to the two holons that propagate with the same velocity but different momenta and do not scatter off each other in the integrable model. The existence of this divergent edge at finite $U$, near the zone boundary  and at finite energies, is confirmed by the tDMRG results. Within the precision of the numerical results, we   found no evidence for rounding of this singularity due to coupling to continuum below it, which would be apparent in the form of an exponential decay of the real-time correlation function.

The agreement between the analytical predictions and the numerical line shapes obtained by tDMRG allowed us to explain how the line shape of $S(q,\omega)$ changes as a function of $U$, interpolating between the strong coupling  and the weak coupling limits. Starting from strong coupling and decreasing $U$, we observed that the spectral weight inside the two-holon continuum decreases while the spectral weight below the lower edge of the two-holon continuum increases. The $U\to 0$ limit is nonperturbative, as expected from spin-charge separation and the Mott transition, and this is manifested in the dynamic response function through a discontinuous change in the edge exponents. For instance, while at $U=0$ $S(q,\omega)$ has a square-root divergence below the upper threshold of the electron-hole continuum, for arbitrarily small $U$ this singularity is removed and a square-root divergence forms above the lower threshold of the two-holon continuum. 

We end by commenting on the connection with experiments that show a sharp feature observed in the RIXS spectrum of 1D Mott insulators for momentum near the zone boundary.\cite{rixs,gu} This feature was interpreted as an exciton in Ref. \onlinecite{gu}, expected from the strong coupling theory  for the extended Hubbard model, but as a broad two-holon resonance in Ref. \onlinecite{rixs}. Our results for $S(q,\omega)$ of  the integrable Hubbard model do not have any excitonic bound states, but also show a sharp feature near the zone boundary which is actually a  square-root divergence at the lower edge of the two-holon continuum at finite $U$. Therefore, a possible interpretation of the experiments is that the sharp feature is the result of a slight rounding of  this Van Hove singularity in a system where the integrability breaking interactions (primarily the nearest neighbour interaction in the extended Hubbard model) are fairly weak. However, the   nearest neighbor interaction is not guaranteed to be negligible since screening is typically rather weak in insulators such as Sr2 CuO3 .

\acknowledgements

We thank I. Affleck, F.  Essler and  A. Muramatsu for illuminating discussions. This research was supported  by  the Brazilian CNPq grant 309234/2011-5, the FCT Portuguese grant PTDC/FIS/64926/2006, the NSF under DMR 090-7500, German transregional collaborative research center SFB/TRR21, and Max Planck Institute for Solid State Research.  JMPC thanks the hospitality of the Institut f\"ur Theoretische Physik III, 
Universit\"at Stuttgart, where part of this research was performed.

\appendix

\section{Symmetry and elementary excitations in the Bethe ansatz solution\label{app:BAnotation}}

Here we briefly discuss the relation of the operational representation of Ref. \onlinecite{companion}
to the excitations considered in this paper.

The pseudofermion dynamical theory\cite{carmelo} employs a unitary transformation originally devised to work in the strong coupling limit\cite{harris}  that rotates electron operators to a basis where double occupancy is a good quantum number. The rotated-electron configurations are then naturally expressed in terms of pseudoparticles whose
discrete momentum values are BA exact quantum numbers. The occupancy configurations of the {\it spin-$1/2$ spinons},
{\it $\eta$-spin-$1/2$ $\eta$-spinons}, and spin-less and $\eta$-spin-less {\it $c$ fermions} of
that representation generate both the representations of the spin $SU(2)$
symmetry, $\eta$-spin $SU(2)$ symmetry, and charge hidden $U(1)$ symmetry algebras,
respectively, and the model $4^{L}$ energy eigenstates. The {\it spin-$1/2$ spinons}
are the spins carried by the rotated electrons of the singly occupied sites.
The {\it $\eta$-spin-$1/2$ $\eta$-spinons} of projection $-1/2$ and $+1/2$ refer
to the $\eta$-spin degrees of freedom of the rotated-electron doubly occupied
and unoccupied sites, respectively. The $c$ fermions describe the charge
hidden $U(1)$ symmetry degrees of freedom of the rotated electrons of the singly occupied sites.
The $c$ fermion holes describe the hidden $U(1)$ symmetry degrees of freedom of the
rotated-electron doubly occupied and unoccupied sites.

The occupancy configurations of the spin-neutral composite $s\nu$ fermions, 
each containing $2\nu$ bound spinons, considered in Ref. \onlinecite{companion}, were
called distributions of magnon bound states by M. Takahashi.\cite{takahashi} Furthermore, the occupancy configurations of 
the $\eta$-spin neutral composite $\eta\nu$ fermions of Ref. \onlinecite{companion}, 
each containing $2\nu$ anti-bound $\eta$-spinons, correspond to his distributions of bound states of pairs.
Specifically, the momentum occupancy configurations of the $c$ fermions, 
$\eta$-spin-neutral $2\nu$-$\eta$-spinon composite $\eta\nu$ fermions, and spin-neutral 
$2\nu$-spinon composite $s1$ fermions where $\nu=1,...,\infty$
generate excitations described by the BA thermodynamic equations (2.12a), (2.12b), and (2.12c) of Ref. \onlinecite{takahashi},
respectively. In units of $2\pi/N_a$, the momentum values of those objects are the BA 
quantum numbers $I_j$, ${J_{\alpha}'}^n$, and $J_{\alpha}^n$
in such equations, respectively. Here within the Ref. \cite{companion} notation, the index
$n=\nu=1,...,\infty$ in  ${J_{\alpha}'}^n$ and $J_{\alpha}^n$ refers to the number of 
anti-bound-$\eta$-spinon pairs and bound-spinon pairs, respectively, and $\alpha =j$
is the momentum value index. 

Note that the two sets of BA thermodynamic equations given in Eqs. (2.12b), and (2.12c) of Ref. \onlinecite{takahashi},
which are associated with $\eta$-spin-singlet and spin-singlet excitations, respectively,
have exactly the same structure. This is consistent with the excitations described by
the BA thermodynamic equation (2.12a) of that reference referring to a degree of freedom
other than $\eta$-spin and spin. Consistently, in Ref. \onlinecite{companion} it is confirmed
that the latter excitations generate representations of the hidden $U(1)$ symmetry
in the model extended global $[SO(4)\times U(1)]/Z_2=[SU(2)\times SU(2)\times U(1)]/Z_2^2$
symmetry.

For the problem studied in this paper, only excitations generated by $c$ momentum band and 
spin-neutral $\nu=1$ two-spinon $s1$ fermion band occupancy configurations play an active role.
Those excitations also contain two $\eta$-spinons, whose occupancies generate the three $\eta$-spin-triplet states.  
The spin-singlet excitations generated by the two-spinon $s1$ fermion momentum occupancy
configurations are described by the BA thermodynamic equations (2.12b) of Ref. \onlinecite{takahashi} for
$n=1$ spinon pairs.

In this paper we call holons and spinons the holes of the $c$ fermion
and $s1$ fermion momentum bands, respectively. Hence the spinons considered here are
spin-neutral objects. This is in contrast to those of Ref. \onlinecite{companion}, which carry spin $1/2$.

In the thermodynamic limit holons and spinons have dispersion relations $\varepsilon_{c}(p_{c})$ and $\varepsilon_s(p_s)$, respectively, where the dressed momenta $p_{c,s}$ and dressed energies $\varepsilon_{c,s}$ are given by,
\begin{equation}
\varepsilon_c (p_c) = {U\over 2} - \epsilon^0_c (q)\vert_{q={\pi\over 2} -p_c}
\, ; \hspace{0.35cm}
\varepsilon_s (p_s) = - \epsilon^0_s (p)\vert_{p={\pi\over 2} -p_s} \, .
\label{dressed-en}
\end{equation}
The energy bands $\epsilon^0_c (q)$ and $ \epsilon^0_s (p)$ and corresponding momenta $q$ and $p$ are given in Eqs. (A1)-(A4) of Ref. \onlinecite{Carmelo92}. For the present half filling case, the relation to Bessel functions provided in Eq. (A8) of that reference applies.

\section{Marginal coupling between chiral spin currents and spin impurity\label{App:RG}}

Consider the marginal operator in Eq. (\ref{intRLD}). In this appendix we derive the renormalization group (RG) equations for this perturbation to the free Hamiltonian in Eq. (\ref{H0invariant}). The  RG with high energy impurity modes is not standard, but the meaning is to investigate the effects of the perturbation when we approach the threshold where Hamiltonian (\ref{H0invariant}) predicts a power-law singularity. The intuitive picture is that, as we approach the threshold, the energy of   particle-hole excitations that the mobile impurity is allowed to scatter is reduced. Therefore, we shall consider the renormalization of the coupling constants $\kappa_{R,L}$ when we integrate out  an energy shell in the sub-bands near the Fermi surface. For consistency, the band width of the impurity modes must be reduced as well, but this effect will not be crucial for our conclusions. 

Let us focus on $\kappa_L$ (the calculation for $\kappa_R$ is completely analogous). We apply the perturbative RG.\cite{cardy} The partition function has the form \be
Z=\mathrm{Tr }\exp\left[-\int d^2x\,(\mc H+\delta\mc H_{RLD})\right].
\ee
Expanding for small $\kappa_L$ (and omitting normal ordering signs), we obtain\bea
Z&\approx& Z_0\left[ 1+2\pi v_s\kappa_L\int d^2x\, \mathbf J_L(x)\cdot  D^\dagger_s(x)\bm\tau D_s^{\phantom\dagger}(x)\right.\nonumber\\
&&+\frac{(2\pi v_s\kappa_L)^2}2\int d^2x\int d^2x^\prime\,  (\tau^a)_{i,j}(\tau^b)_{l,m}\label{pertRG}\\
&&\left. \times  J_L^a(x)J_L^b(x^\prime) D^\dagger_{s,i}(x) D_{s,j}^{\phantom\dagger}(x)D^\dagger_{s,l}(x^\prime) D_{s,m}^{\phantom\dagger}(x^\prime)\right]\nonumber,
\eea
where $Z_0$ is the free part associated with Hamiltonian  (\ref{H0invariant}). The $\mc{O}(\kappa_L^2)$ term can generate corrections to $\kappa_L$ when we integrate out ``fast''  modes. We use the operator product expansion of the spin currents\cite{tsvelik}\be
J^a_L(z)J^b_L(z^\prime)\sim \frac{\delta^{ab}}{8\pi^2z^2}+\frac{i}{2\pi z}\epsilon^{abc}J_L^c(z^\prime)+\dots,\label{opeJ}
\ee
where $z=v_s\tau+ix$ is the complex argument of holomorphic functions. In Eq. (\ref{pertRG}), we must also take contraction of $D_s$ fields. For this purpose we need the impurity propagator in imaginary time\bea
&&\langle T_\tau D^{\phantom\dagger}_{s,i}(x,\tau)D^\dagger_{s,j}(0,0)\rangle\nonumber\\
&=&\delta_{i,j}\theta(\tau)e^{-\varepsilon_s\tau}\int_{-\mc{K}}^{\mc{K}} \frac{dp}{2\pi}\, e^{-p(u_s \tau-ix)}\nonumber\\
&\equiv&\delta_{i,j}\theta(\tau)G(x,\tau),
\eea
where $\mc{K}$ is the momentum cutoff of the impurity sub-band. We obtain\be
G(x,\tau)=e^{-\varepsilon_s\tau}\frac{\sinh[\mc{K}(u_s\tau-ix)]}{\pi(u_s\tau-ix)}.\label{illdefined}
\ee
Note that we cannot take the limit $\mc{K}\to \infty$  in Eq. (\ref{illdefined}) yet. (For the propagator in real time, this is possible and yields the delta function in Eq. (\ref{propagD}).) 

Using Eqs. (\ref{opeJ}) and (\ref{illdefined}) in Eq. (\ref{pertRG}), we find (keeping only corrections to $\kappa_L$)\bea
Z&\approx& Z_0\left[1+2\pi v_s\kappa_L\int d^2x\, \mathbf J_L\cdot D^\dagger_s\bm\tau D_s^{\phantom\dagger}\right.\nonumber\\
&&-\pi (v_s\kappa_L)^2\int d^2x \, \mathbf J_L\cdot D^\dagger_s\bm \tau D_{s}^{\phantom\dagger}\nonumber\\
&& \left.\times \int d^2\tilde {x}\, \frac{ \textrm{sign} (\tilde\tau)}{v_s\tilde\tau+i\tilde x}\,G(\tilde x,\tilde\tau)\right],\label{zint2}
\eea
where $(\tilde\tau,\tilde x)=(\tau-\tau^\prime,x-x^\prime)$ are the relative coordinates of the two points in Euclidean space-time.  Importantly, the impurity propagates with a different velocity than the bosonic modes, thus the problem is not Lorentz invariant. Physically, this is more like a boundary problem, with a ``mobile boundary'' represented by the impurity that the bosonic modes have to track. Therefore, instead of a rotationally symmetric energy-momentum shell, we integrate out the ``fast'' modes   contained in the  strip $-\infty<\tilde x<\infty$, $1/\Lambda<|\tilde\tau|<1/\Lambda^\prime$, with $\Lambda$ and $\Lambda^\prime=\Lambda-d\Lambda$ being the original and reduced energy cutoffs, respectively. The integration over $\tilde x$ gives\be
\int_{-\infty}^\infty d\tilde {x}\, \frac{G(\tilde x,\tilde\tau)}{v_s\tilde\tau+i\tilde x}=e^{-\varepsilon_s\tilde\tau}\frac{1-e^{-\mc{K}(v_s+u_s) |\tilde\tau|}}{(v_s+u_s)\tilde\tau}.\label{dropcutoff}
\ee
We can take the limit $\mc{K}\to\infty$ in Eq. (\ref{dropcutoff}). Moreover, we are integrating out short time differences $\tilde\tau\ll 1/\Lambda\sim 1/\varepsilon_s$, thus we can approximate $e^{-\varepsilon_s\tilde\tau}\approx 1$. We are left with the imaginary time integral\be
\frac2{v_s+u_s}\int_{1/\Lambda}^{1/\Lambda^\prime}\frac{d\tilde\tau}{\tilde\tau}=\frac{2d\ell}{v_s+u_s},
\ee
where $d\ell=d\Lambda/\Lambda$. 

Finally, substituting the result in Eq. (\ref{zint2}) and reexponentiating, we find the RG equation for $\kappa_L$:\be
\frac{d\kappa_L}{d\ell}=-\frac{v_s}{v_s+u_s}\kappa_L^2.\label{RGkappaL}
\ee
The RG equation for $\kappa_R$ is obtained from Eq. (\ref{RGkappaL}) by the substitution $\kappa_L\to\kappa_R,u_s\to-u_s$. Since $u_s<v_s$, we conclude that $\kappa_L>0$ and $\kappa_R>0$ are marginally irrelevant. We believe this to be the correct sign for the coupling constants of the Hubbard model. Furthermore, we expect  the marginally irrelevant $\kappa_{L,R}$ operators to give rise to logarithmic corrections to   edge singularities for $SU(2)$ symmetric models, similarly to the effect in equal-time correlation functions.\cite{gepner}  Logarithmic corrections are known to exist at the lower edge of  the two-spinon contribution to the spin DSF for the Heisenberg model,\cite{karbach} but we do not pursue that calculation here.

\section{Exponent for threshold with  two charge impurities and   one spin impurity\label{app:exp32}}

In this appendix we detail the calculation of the exponent for  the threshold $\omega_{2c2s}(q_\triangleleft<q<q_\triangleright)$ in $S(q,\omega)$, which is described by two high energy holons and one high energy spinon,  all moving with the same velocity.  Other exponents can be obtained by similar methods. 

We find it convenient to calculate the exponent of $S(q,\omega)$ using the analytical continuation of  imaginary time propagators to real time prescribed as follows.  The zero temperature limit of the imaginary time propagator of the density operator is\be
\langle n_q(\tau)n_{-q}(0)\rangle=\sum_\nu|\langle GS|n_q|\nu\rangle|^2 e^{-(E_\nu-E_{GS})\tau}.
\ee
Using the analytic continuation with the prescription $i\tau\to -(t-i\eta)$, we obtain
\be
\langle n_q(t-i\eta)n_{-q}(0)\rangle=\sum_\nu|\langle GS|n_q|\nu\rangle|^2 e^{-i(E_\nu-E_{GS})(t-i\eta)},
\ee
where $\eta\to 0^+$ at the end guarantees the convergence of the sum. Taking the Fourier transform, we get\bea
&&\int_{-\infty}^{+\infty} dt\,e^{i\omega t }\langle n_q(t-i\eta)n_{-q}(0)\rangle\nonumber\\
&=&2\pi\sum_\nu|\langle GS|n_q|\nu\rangle|^2 \delta(\omega-E_\nu+E_{GS}),
\eea
which is the correct expression for $S(q,\omega)$.

As argued in Sec. \ref{sec:edgetheory}, due to  $\eta$-spin $SU(2)$ symmetry the exponent for $\eta^z=0$ excitation is the same as the exponent for the $\eta^z=+1$ excitation. In the latter case we can treat  the two holons  as identical  spinless fermions that do not interact via $s$-wave scattering. The only interaction in this three-body problem is between the holons and the spinon. In first quantization, we    write down the effective Hamiltonian (for energies measured from the threshold)
\bea
H_{3b}&=&\frac{p_{1}^2+p_{2}^2}{2m_c}+\frac{p_{3}^2}{2m_s}+u(p_1+p_2+p_3)\nonumber\\
&&+V_{cs}[\delta(x_1-x_3)+\delta(x_2-x_3)],\label{H3beff}
\eea
where particles 1 and 2 are the two holons and particle 3 is the spinon, with canonically conjugated variables $[x_n,p_m]=i\delta_{nm}$. For a generic spinon-holon interaction potential, the parameter $V_{cs}$ is related to the $s$-wave scattering length. The wave functions in the physical Hilbert space must be anti-symmetric with respect to exchanging 1 and 2. The three-body propagator in imaginary time  can be calculated from \be
G_{3b}(x,\tau)=\langle\Phi|e^{iPx}e^{-H_{3b}\tau}|\Phi\rangle,
\ee
where $P=p_1+p_2+p_3$ is the total momentum operator and \bea
|\Phi\rangle&=&\frac1{\sqrt2}\left( \left|x_1=\frac \varepsilon2,x_2= -\frac\varepsilon2\right\rangle- \left|x_1=-\frac\varepsilon2,x_2= \frac\varepsilon2\right\rangle\right)\nonumber\\
&&\otimes|x_3=0\rangle\label{Psi3b}
\eea
is the initial state created by applying  $d_c^\dagger(x+\frac\varepsilon2)d^\dagger_c(x-\frac\varepsilon2)d_s^\dagger(0)$ on the ground state. 

We perform a change of variables from $x_1,x_2,x_3$ to $X=[m_c(x_1+x_2)+m_sx_3]/(2m_c+m_s), x_r=x_1-x_2, z=x_1+x_2-2x_3$ and the associated conjugate  momenta. The Hamiltonian becomes\bea
H_{3b}&=&\frac{P^2}{2(2m_c+m_s)} +uP+\frac{p_{r}^2}{2m_r}+\frac{p_{z}^2}{2m_z}\nonumber\\
&&+2V_{cs}[\delta(x_r-z)+\delta(x_r+z)],\label{H3b2}
\eea
where $m_r=m_c/2$ and $m_z=\frac{m_cm_s(m_s+2m_c)}{2(2m_s+m_c)^2}$. In terms of these new variables, the initial state has $X=z=0$, $x_r=\varepsilon$. The $x$ dependence of $G_{3b}(x)$ is entirely in the free centre-of-mass ``particle''.  
 We note that $m_r\neq m_z $ $\forall m_c,m_s\in\mathbb R$. While $m_r>0$ for holons below the inflection point, we   assume $m_z>0$ as well, which is easily verified in the strong coupling limit. 

First consider the simpler case $V_{cs}=0$. In this case, all three particles are free and the propagator factorizes\be
G_{3b}(x,\tau)=G_{cm}(x,\tau)G_r(\tau)G_z(\tau).
\ee
For the propagator of the centre of mass particle, which moves with velocity $u$, we shall use as in Eq. (\ref{illdefined})\be
G_{cm}(x,\tau)=\frac{\sinh[\mc K(u\tau-ix)]}{\pi(u\tau-ix)},
\ee
with cutoff $\mc{K}\ll |(2m_c+m_s)u|$. For the other ``particles'', we have \be
G_{r}(\tau)\sim \int_{-\infty}^{\infty} dk_r\,\sin^2(k_r\varepsilon)e^{-k_r^2\tau /2m_r}\sim \varepsilon^2  \left(\frac{m_r}\tau\right)^{3/2},\label{Gr32}
\ee
for $\tau\gg |m_r|\varepsilon^2$ and
\be
G_{z}(\tau)\sim \int_{-\infty}^{\infty} dk_z\,e^{-k_z^2\tau /2m_z}\sim \left(\frac{m_z}\tau\right)^{1/2}.
\ee
Notice that $G_r(\tau)$ decays faster because of fermionic statistics, which imposes that the wave function is an odd function of $x_r$.  This is equivalent to the vanishing matrix element in Ref. \onlinecite{penc00}. 

At the  threshold $\omega_{2c2s}(q_\triangleleft<q<q_\triangleright)$, the three body propagator has to be combined with the low energy propagator of the chiral spinor, which has scaling dimension 1/4. The integral over $x$ gives (for $v_s>u$, there are two separate contributions from a pole and a branch cut in the lower half plane)\be
\int_{-\infty}^{\infty}dx\,\frac{G_{cm}(x,\tau)}{(v_s\tau-ix)^{1/2}}\sim \frac{1}{[(v_s-u)\tau]^{1/2}},
\ee
in which we took  the limit $\mc K\to \infty$ after the integration. 
Combining with $G_r(\tau)$ and $G_z(\tau)$ and switching to real time $i\tau\to -(t-i\eta)$ as explained above, the remaining time integral gives\be
S(q,\omega)\sim\int_{-\infty}^{\infty} dt\frac{e^{i\delta\omega t}}{(t-i\eta)^{5/2}}\sim\theta(\delta\omega)\delta\omega^{3/2}.\label{expccs}
\ee

Now consider $V_{cs}\neq 0$. In this case the $x_r$ and $z$ particles are scattered by the potentials in Eq. (\ref{H3b2}). Nonetheless, we argue that the edge exponent is the same as for $V_{cs}=0$. First, we note that the exponent depends on the long-time behavior of $G_{3b}(x,t)$, which in turn depends on the behavior of low energy  eigenfunctions   for $z=0$, $x_r=\varepsilon\to0$. The extra power of $1/\tau$ in Eq. (\ref{Gr32}) is a result of the wave function vanishing as $\sim k_r\varepsilon$ for $k_r\varepsilon\to0. $ Then we must ask whether $V_{cs}\neq0$ modifies the behavior of the wave function in the long wavelength limit. 

Rescaling $z\to  z\tan\alpha$, $p_z\to  p_z\cot\alpha$ with $\tan\alpha=\sqrt{m_r/m_z}\neq1$ in Eq. (\ref{H3b2}), the $x_r$ and $z$ part of the Hamiltonian becomes\be
H_{2b}=\frac{p_r^2+p_z^2}{2m_r}+2V_{cs}\delta(x_r\pm z\tan\alpha ).
\ee
We then introduce polar coordinates $z=\rho\cos\phi,x_r=\rho\sin\phi$. The two-dimensional Schr\"odinger equation for the wave function $\Phi(x_r,z)=\Phi(\rho,\phi)$ reads\bea
&&\frac1\rho\frac{\partial}{\partial \rho}\left(\rho\frac{\partial\Phi}{\partial \rho}\right)+\frac1{\rho^2}\frac{\partial^2\Phi}{\partial \phi^2}+k^2\Phi\nonumber\\
&=&\frac{2m_rV_{cs}\cos\alpha}{\rho}[\delta(\phi\pm\alpha)+\delta(\phi\pm\alpha+\pi)],\label{schrod}
\eea
where $k$ is related to the energy  by  $E=k^2/2m_r$.

Equation (\ref{schrod}) describes the motion of a particle in two dimensions which is scattered by delta function potentials located along  the lines $x_r=\pm z\cot\alpha$. We can solve the wave functions in the four regions of the $(z,x_r)$ plane separated by these lines and then match the wave functions with a   discontinuity in $\partial\Phi/\partial\phi$ at the boundaries. The solutions are of the form $\Phi(\rho,\phi)=\sum_{n=0}^\infty[A_ne^{in\phi}+B_ne^{-in\phi}]J_n(k\rho)$, where $J_n$ denotes the Bessel function of the first kind. Imposing that the wave function is continuous everywhere and  is  anti-symmetric with respect to exchanging the two holons implies that $\Phi(\rho=0,\phi)=0$, hence  $A_0=B_0=0$ in all regions.

In the long wavelength limit, $k\to0$, the delta function potentials  become impenetrable and the wave function vanishes along the lines $x_r=\pm z\cot\alpha$. Importantly, these lines do not coincide with the $z,x_r$ axis (in which the kinetic energy is diagonal) since $m_r\neq m_z$. But we are interested in the behavior of the wave function for $z=0$, $x_r=\varepsilon\to0$, i.e. approaching the origin along the $x_r$ axis. The wave function already vanishes at $\rho=0$ due to the anti-symmetrization, therefore it is not affected by the delta function potential at $x_r=z=0$. As a result,  for $k\varepsilon\to0$ the eigenfunctions vanish as $\Phi(\rho=\varepsilon,\phi=\pi/2 )\sim J_{1}(k\varepsilon)\sim k\varepsilon$. This is the same behavior as obtained for $V_{cs}=0$ and leads to the   exponent $\mu=3/2$ in $S(q,\omega)$ as in Eq. (\ref{expccs}).

\section{Absence of scattering between distinguishable charge impurities  in the integrable model \label{integrable}}

In this appendix we show that the coupling constants $V_c^{C}$ and $V_c^E$ in Eq. (\ref{CandE}) are fine tuned to zero as a result of the integrability of the Hubbard model. Here integrability is understood as the existence of an infinite number of local conserved quantities   in the thermodynamic limit. The simplest  nontrivial conserved quantity of the Hubbard model was discovered by Shastry\cite{shastry} and can be written as\cite{zotos}\bea
Q_3&=&\sum_{j,\sigma}\left[(ic^\dagger_{j+1,\sigma}c_{j-1,\sigma}+\textrm{h.c.})\right.\nonumber\\
&&\left. - U( \mc J_{j-1,\sigma}+\mc J_{j,\sigma})\left(n_{j,-\sigma}-\frac12\right)\right],
\eea
where $\mc J_{j,\sigma}=ic^\dagger_{j+1,\sigma}c_{j,\sigma}+\textrm{h.c.}$ is the current density operator for electrons with spin $\sigma$. The conserved quantity  $Q_3$ is \emph{almost} equal to  the energy current operator, differing only by a factor of 2 in front of $U$.\cite{zotos} 
The   energy current operator  $\mc J^E=\sum_j  \mc J^E_j$ is defined from the continuity equation of the Hamiltonian density. Writing $H=\sum_jh_j$ with \be
h_j= -(c^\dagger_{j}c^{\phantom\dagger }_{j+1}+\textrm{h.c.})+U\left(n_{j,\uparrow}-\frac12\right)\left(n_{j,\downarrow}-\frac12\right),
\ee
we obtain $ \mc J^E_j$ by taking the commutator of $h_j$ with $H$, which   has the form of a discretized divergence\be
i[h_j,H]= \mc J^E_{j+1}- \mc J^E_j.\label{continuityH}
\ee 
The operator $Q_3$ can be written as\be
Q_3=2\mc J^E+Y,\label{q3trick}
\ee
where $Y=\sum_j Y_j= \sum_j(-i c^\dagger_{j+1}c^{\phantom\dagger }_{j-1}+\textrm{h.c.})$. Interestingly,   $Y$ is independent of $U$ and its density appears in the  the commutator of the charge current density $\mc J_j=\sum_\sigma \mc J_{j,\sigma}$ with the total charge current $\mc J=\sum_{j} \mc J_{j}$:\be
-i[\mc J_j,\mc J]=Y_{j+1}-Y_j.\label{commutJ}
\ee
 
We want to impose the conservation of $Q_3$ in the effective model Eq. (\ref{CandE}). A  similar idea has been applied to the XXZ model,\cite{pereiraJSTAT} in which case  it was shown that   conservation laws lead to constraints on   irrelevant operators at low energies, with consequences for dynamic correlation functions.  Since the impurity model is phenomenological, we need a prescription to construct the conserved quantity directly in the field theory. The key is to use the continuity equations and relations (\ref{q3trick}) and (\ref{commutJ}) since currents can be easily  identified in the field theory. A caveat in applying Eq. (\ref{commutJ}) in the field theory is that the dimensions of  the density of $Y$ and $\mc J^E$ differ by a factor of lattice spacing squared. This entails that when combining $Y$ from Eq. (\ref{commutJ}) with $\mc J^E$  from Eq. (\ref{continuityH}) we must restore   nonuniversal  factors of short distance cutoff for dimensional analysis.

The calculation of $\mc J^E$ and $Y$ in the field theory   can be simplified using the  local $SU(2)$ algebra of $ D^\dagger_c\bm\tau D_c^{\phantom\dagger}$ and  $ \tilde D^\dagger_c\bm\tau \tilde D_c^{\phantom\dagger}$. The charge current density obtained from the continuity equation for the charge density $n(x)\sim D^\dagger_c(x)\tau^zD_c^{\phantom\dagger}(x)+\tilde D^\dagger_c(x)\tau^z\tilde D_c^{\phantom\dagger}(x)$ is\bea
\mc J(x)&=&D^\dagger_c(x)\tau^z\left(u_c-\frac{i}{m_c}\partial_x\right)D_c^{\phantom\dagger}(x)\nonumber\\
&&+\tilde D^\dagger_c(x)\tau^z\left(u_c-\frac{i}{\tilde m_c}\partial_x\right)\tilde D_c^{\phantom\dagger}(x).
\eea
The commutator of the charge current density with the integrated charge current $\mc J=\int dx^\prime \mc J(x^\prime)$ gives\bea
i[\mc J(x),\mc J]&=&\partial_x\left[D^\dagger_c(x)\left(\frac{u_c}{m_c}-\frac{i}{m_c^2}\partial_x\right)D_c^{\phantom\dagger}(x)\right.\nonumber\\
&&+\left.\tilde D^\dagger_c(x)\left(\frac{u_c}{\tilde m_c}-\frac{i}{\tilde m_c^2}\partial_x\right)\tilde D_c^{\phantom\dagger}(x)\right].\quad 
\eea
Comparing with Eq. (\ref{commutJ}), we conclude that the continuum version of $Y$ is $Y=\int dx\,Y(x)$ with density \bea
Y(x)&=&-D^\dagger_c(x)\left(\frac{u_c}{m_c}-\frac{i}{m_c^2}\partial_x\right)D_c^{\phantom\dagger}(x)\nonumber\\
&&-\tilde D^\dagger_c(x)\left(\frac{u_c}{\tilde m_c}-\frac{i}{\tilde m_c^2}\partial_x\right)\tilde D_c^{\phantom\dagger}(x).
\eea

The energy current operator   is obtained from the commutator $i[\mc H(x),\int dx^\prime \, \mc H(x^\prime)]=\partial_x\mc J^E(x)$. We find\bea
\mc J^E(x)&=&\varepsilon_cD^\dagger_c\left(u_c-\frac{i}{m_c}\partial_x\right)D_c^{\phantom\dagger}\nonumber\\
&&+\tilde\varepsilon_c \tilde D^\dagger_c\left(u_c-\frac{i}{\tilde m_c}\partial_x\right)\tilde D_c^{\phantom\dagger}\nonumber\\
&&+2u_cV_c^CD^\dagger_cD_c^{\phantom\dagger}\tilde D^\dagger_c\tilde D_c^{\phantom\dagger}\nonumber\\
&&+2u_cV_c^ED^\dagger_c\bm\tau D_c^{\phantom\dagger}\cdot \tilde D^\dagger_c \bm \tau\tilde D_c^{\phantom\dagger},
\eea
where we neglect operators with dimension higher than 2.

Using Eq. (\ref{q3trick}), we construct the density of the conserved quantity\bea
Q_3(x)&=&2E_c D^\dagger_c\left(u_c-\frac{i}{m_c}\partial_x\right)D_c^{\phantom\dagger}\nonumber\\
&&+2 \tilde E_c\tilde D^\dagger_c\left(u_c-\frac{i}{\tilde m_c}\partial_x\right)\tilde D_c^{\phantom\dagger}\nonumber\\
&&+4u_cV_c^CD^\dagger_cD_c^{\phantom\dagger}\tilde D^\dagger_c\tilde D_c^{\phantom\dagger}\nonumber\\
&&+4u_cV_c^ED^\dagger_c\bm\tau D_c^{\phantom\dagger}\cdot \tilde D^\dagger_c \bm \tau\tilde D_c^{\phantom\dagger},\label{Q3ft}
\eea
where $E_c=\varepsilon_c-1/(2m_c\alpha^2)$, $\tilde E_c=\tilde \varepsilon_c-1/(2\tilde m_c\alpha^2)$, with $\alpha$ the short distance cutoff. The density  of $Q_3$ in Eq. (\ref{Q3ft}) contains all the operators up to dimension 2 that are invariant under $\eta$-spin rotation but with different coefficients than the Hamiltonian (\ref{CandE}). In fact,  $Q_3$ has  the same symmetries as the Hamiltonian except for the signature under parity transformation (parity symmetry is broken by hand in the effective impurity model by the definition of the impurity sub-bands).

Finally, taking the commutator of   $Q_3=\int dx\, Q_3(x)$ with $H=\int dx\; \mc H(x)$, we are left with  two dimension-three operators that do not not vanish  in general\bea
[ Q_3,H]&=&2i\left(\frac{E_c}{m_c}-\frac{\tilde E_c}{\tilde m_c}\right)\left[V^C_c\int dx\,\partial_x(D^\dagger_cD^{\phantom\dagger}_c)\tilde D^\dagger_c\tilde D^{\phantom\dagger}_c\right.\nonumber\\
&&+\left.V^E_c\int dx\,\partial_x(D^\dagger_c\bm \tau D^{\phantom\dagger}_c)\cdot \tilde D^\dagger_c\bm\tau \tilde D^{\phantom\dagger}_c\right].\label{JeH}
\eea
We note that other terms cancel because the two sub-bands have the same velocity $u_c$. Recall that along the boundary line $\omega_{2c}^-(q>q_\diamond)$ we have $m_c>0$ and $\tilde m_c<0$ since   the $D_c$ sub-band is  below the inflection point of  the holon dispersion and the $\tilde D_c$ sub-band is above it.  Moreover, $m_c>|\tilde m_c|$ because  the curvature of the holon dispersion is smaller close to the band minimum. Thus we have\be 
\frac{E}{m_c}-\frac{\tilde E_c}{\tilde m_c}=\frac{\varepsilon_c}{m_c}+\frac{\tilde \varepsilon_c}{|\tilde m_c|}+\frac{1}{2|\tilde m_c|\alpha^2}-\frac{1}{ 2m_c\alpha^2}>0.
\ee

The only way to ensure that the commutator in Eq. (\ref{JeH}) vanishes  is to set $V_c^C=V_c^E=0$. Therefore, the existence of a conserved quantity represented in the field theory by an operator of  the form in Eq. (\ref{Q3ft}) requires that there is no scattering between $D_c$ and $\tilde D_c$ holons. Importantly, integrability does not have any implications for the interaction   between two impurities within the same spinor ($V_s$ in Eq. (\ref{Hexciton}) and $V_c$ in Eq. (\ref{H2clower})). This follows from taking $\tilde\varepsilon_c= \varepsilon_c$ and $\tilde m_c= m_c$ in Eq. (\ref{JeH}), in which case the commutator  vanishes identically.

We also remark that the effective model  in principle also contains irrelevant   interactions that have the same dimension (three) as the parabolic dispersion term. These irrelevant interactions, which  were omitted in Eq. (\ref{CandE}),  can contribute to the coefficient of the last two terms in the conserved quantity in Eq. (\ref{Q3ft}). However, such  terms do not contribute to the commutator in Eq. (\ref{JeH}) (at the level of dimension-three operators), thus our conclusion is not affected by irrelevant interactions.


\begin{thebibliography}{99}
\bibitem{foot1}In terms of BA quantum numbers (as defined in Takahashi's equations, Ref. \onlinecite{takahashi}), the lowest-weight state in the charge triplet excitation that corresponds to this energy has quantum numbers $\{J_\alpha\}=\{-3/2,-1/2,1/2,3/2\}$   and $\{I_j\}=\{-4,-3,-2,-1,1,2,3,4\}$.
\end{thebibliography}
\end{document}